\documentclass[preprint2,numberedappendix]{emulateapj}

\usepackage{amsmath,amssymb,latexsym,graphics,epsfig,exscale,graphicx}
\usepackage{subfigure}
\usepackage{pstricks}
\usepackage{graphicx}
\usepackage{verbatim}
\usepackage{rotating}

\slugcomment{Submitted to ApJ.}
\shorttitle{{\it Single parameter galaxy classification}}
\shortauthors{M. Taghizadeh-Popp et al.}

\begin{document}

\title{Single parameter galaxy classification: The Principal Curve through the multi-dimensional space of galaxy properties}

\author{M. Taghizadeh-Popp \altaffilmark{1}, S. Heinis\altaffilmark{1,2}, A. S. Szalay \altaffilmark{1}}
\altaffiltext{1}{Department of Physics and Astronomy, Johns Hopkins University. 3400 North Charles Street, Baltimore, MD 21218, USA. e-mail: mtaghiza [at] pha.jhu.edu}
\altaffiltext{2}{Laboratoire d'Astrophysique de Marseille, OAMP, Universit\'e Aix-marseille, CNRS, 38 rue Fr\'ed\'eric Joliot-Curie, 13388 Marseille cedex 13, France}

\begin{abstract}

We propose to describe the variety of galaxies from the Sloan Digital Sky Survey (SDSS) by using only one affine parameter. To this aim, we construct the Principal Curve (P-curve) passing through the spine of the data point cloud, considering the eigenspace derived from Principal Component Analysis (PCA) of morphological, physical and photometric galaxy properties. Thus, galaxies can be labeled, ranked and classified by a single arc length value of the curve, measured at the unique closest projection of the data points on the P-curve.
We find that the P-curve has a "W" letter shape with 3 turning points, defining 4 branches that represent distinct galaxy populations. This behavior is controlled mainly by two properties, namely $u-r$ and $SFR$ (from blue young at low arc length to red old at high arc length), while most of other properties correlate well with these two.
We further present the variations of several important galaxy properties as a function of arc length. Luminosity functions variate from steep Schechter fits at low arc length, to double power law and ending in Log-normal fits at high arc length. Galaxy clustering shows increasing autocorrelation power at large scales as arc length increases. Cross correlation of galaxies with different arc lengths shows that the probability of 2 galaxies belonging to the same halo decreases as their distance in arc length increases.

PCA analysis allowed to find peculiar galaxy populations located apart from the main cloud of data points, such as small red galaxies dominated by a disk, of relatively high stellar mass-to-light ratio and surface mass density. On the other hand, the P-curve helped understanding the average trends, encoding 75\% of the available information in the data.

The P-curve allows not only dimensionality reduction, but also provides supporting evidence for relevant physical models and scenarios in extragalactic astronomy:

1) Evidence for the hierarchical merging scenario in the formation of a selected group of red massive galaxies. These galaxies present a log-normal r-band luminosity function, which might arise from multiplicative processes involved in this scenario.

2) Connection between the onset of AGN activity and star formation quenching as mentioned in \cite{martin2007}, which appears in green galaxies when transitioning from blue to red populations.

\end{abstract}

\keywords{cosmology: large-scale structure of universe --- galaxies: general  --- galaxies: luminosity function, mass function --- galaxies: fundamental parameters --- galaxies: absorption lines --- 
galaxies: emission lines --- galaxies: photometry --- galaxies: statistics --- methods: statistical --- methods: data analysis  }


\section{Introduction}\label{Sec:Introduction}

In order to constrain the physical processes driving galaxy evolution, it is common practice to measure a number of physical properties for a set of galaxies, and then investigate the correlations between these parameters. In this context, galaxy surveys have become more and more appropriate. The number of galaxies available is getting larger, and the amount of information to constrain physical properties is also increasing, yielding to more accurate estimates. The level of precision of these estimates is also likely to increase in the future, either with the combination of wide angle surveys observing at different wavelengths, or with panchromatic surveys using large number of filters \citep[e.g. PAU,][]{Benitez_2009}, which will benefit from multiband imaging for millions of galaxies. As this data deluge is turning astronomy into becoming a data-intensive or e-science \citep[see ][]{hey2009}, one is confronted with the issue of being just able enough to analyze the feature space, whose dimensionality keeps on increasing. In face of such large amount of physical properties, one wants to find the minimal and most important set which describes galaxies accurately. In this context, a common approach used to reduce the dimensionality of these dataset is performing a Principal Component Analysis \citep[PCA, also known as Karhunen-Lo\`{e}ve transform; see e.g.][]{Efstathiou_1984, Murtagh_1987}. PCA enables us to find an uncorrelated and orthonormal set of linear combinations of properties (eigenvectors) that describe optimally the correlations and variation of the data. This approach has been fruitfully used in astronomy to classify galaxy and quasars based on their spectra \citep{connolly1995,yip2004a, yip2004b}. PCA has be applied on a wider basis using various galaxy properties such as the equivalent width of emission lines \citep{gyory2011} or a mix of spectral and morphological features \citep{coppa2011} to help characterizing the galaxy population. PCA also showed useful for instance when applied to stellar synthesis population models to derive galaxy physical parameters \citep{chen2011}. PCA does not however, enable to capture all the information contained in the input sample. It is by nature linear, and hence can not describe non linear correlations within the data. Other methods, such as applying locally linear embedding to galaxy spectra \citep{Roweis_2000, Vanderplas_2009}, enable to take into account non linearities, as they map high dimension data onto a surface, while preserving the local geometry of the data.

In this paper, we introduce the principal curve \citep[P-curve, see e.g.][for a review]{Einbeck_2007}, which can be seen as a nonparametric extension of linear PCA. The principal curve is the curve following the location of the local mean in the multi-dimensional cloud of data points. In practice, the P-curve can be conveniently built in the PCA eigenspace spanned by the most important eigenvectors along which the variance is highest. The important fact here is that every data point can be assigned a unique closest projection onto the curve, and be labeled by the arc length value measured from the beginning of the curve to the projection. This reduces the complexity of multi dimensional data effectively into only one dimension. Moreover, the ranking of galaxies according to their associated arc length values provides a natural and objective way of ordering, partitioning and classifying the rich zoo of galaxies in the nearby universe.

In this paper, we take advantage of the wealth of data and build the principal curve for both physical and photometric properties belonging to the low redshift Main Galaxy Sample (MGS) \citep{strauss2002}   in SDSS \citep{stoughton2002}. Since the MGS is flux limited, the Malmquist bias underestimates the volume density of faint galaxies compared to that of brighter ones. As a result, the common practice of performing a simple PCA for all galaxies does indeed provide a biased result toward the behavior of the properties of bright objects. As a solution, we do not restrain the statistics by constructing a much smaller volume limited sample, but keep all galaxies by assigning them weights with which we perform Weighted PCA (WPCA) and P-curve methods. We then investigate how the arc length associated to each galaxy correlates with a number of photometric, spectroscopic and physical galaxy properties, as well as morphology, mean spectra, and a first (luminosity function) and second (clustering) moments of galaxies. Our results show that the arc length values remarkably encode a large number of well-known trends in the local Universe.

This paper is organized as follows: in Section \ref{Sec:GalaxySample} we present the dataset we use. Section \ref{Sec:SelectingGalProp} details the galaxy properties we include in our PCA analysis. Section \ref{Sec:MethodDimReduc} presents the methods we use for the dimensionality reduction, weighted PCA and principal curve. We detail in Sec. \ref{Sec:BuildingPcurve} how we build the principal curve from the SDSS data. In Section \ref{Sec:ArcLengthDependance} we present our results and discuss them in Sec. \ref{Sec:Discussion}.

We use in this paper a flat $\Lambda$ cosmology assuming $\lbrace\Omega_{\lambda}$,$\Omega_{\rm M}$,$h_{0}$,$w_{0}\rbrace$ = $\lbrace$0.7,0.3,0.7,-1$\rbrace$.


\section{The Galaxy Sample}\label{Sec:GalaxySample}

In this paper we use photometric and spectroscopic data of galaxies from SDSS-DR8 \citep{aihara2011}, available in a MS-SQL Server database queried online via CasJobs \footnote{ {\tt http://casjobs.sdss.org} }.

In particular, we use the Main Galaxy Sample (MGS) \citep{strauss2002}. These galaxies constitute a flux limited sample, with an r-band petrosian apparent magnitude cut of $m_r \leq 17.77$, and a redshift distribution peaking at $z \sim 0.1$. Their spectra covers the rest frame range of 3800-8000\AA, with a resolution of 69 km s$^{-1}$pix$^{-1}$.

Several selection cuts and flags were enforced in order to have a clean sample. We selected only {\tt science primary} objects appearing in calibrated images having the {\tt photometric} status flag. 
Also, we selected imaging fields where $0.6 \leq {\tt score} \leq 1.0$, which assures good imaging quality with respect to the sky flux and the PSF's width. Furthermore, we neglected individual objects with bad deblending ({with flags \tt PEAKCENTER, DEBLEND$\_$NOPEAK, NOTCHECKED}) and interpolation problems ({\tt PSF$\_$FLUX$\_$INTERP, BAD$\_$COUNTS$\_$ERROR}) or suspicious detections ({\tt SATURATED NOPROFILE}) \footnote{Detailed explanation in \url{\tt sdss3.org/dr8/algorithms}}. Also, we chose galaxies whose spectral line measurements and properties are labeled as {\tt RELIABLE}.

The sky footprint of the clean spectroscopic survey builds up from a complicated geometry defined by {\tt sectors}, whose aggregated area covers $\sim 7930$deg$^{2}$ or a fractional area $F_{A}\simeq 0.192$ of whole sky. We choose a redshift window of $[z_{1},z_{2}]=[0.02,0.08]$. The lower limit avoids including large photometrically-cumbersome galaxies on the sky, and the upper limits reduces the amount of evolution of galaxy properties ($\Delta t < 0.78$Gyr), while keeping the statistics high.
Redshift incompleteness arises from the fact that two 3'' aperture spectroscopic fibers cannot be put together closer than 55'' in the same plate. As an strategy, denser region in the sky are given a greater number of overlapping plates. Nevertheless, 7$\%$ of the initial galaxies photometrically targeted as MGS didn't have their spectra taken.

We further construct a magnitude limited sample, on which we will center our main study. Here, extinction-corrected petrosian apparent magnitude cuts of $[m_{r,1},m_{r,2}]=[13.5,17.65])$ are applied. The lower limit is set due to the arising cross talk from close fibers in the spectrographs, when they contain light from very bright galaxies. The upper limit safely avoids the slight variations of the limiting apparent magnitude around 17.77 over the sky in the targeting algorithm. This leaves us with 174,266 galaxies. 

A volume limited subsample was also created, being a subset of the previous magnitude limited sample. This subsample is used for the study of spatial correlation functions in Sec. \ref{Sec:Clustering}. The redshift ranges are $[z_{1},z_{2}]=[0.02,0.05]$, with an absolute magnitude window of $[M_{r,1},M_{r,2}]=[-21.19,-19.08]$, which leave us with $\sim 40,000$ galaxies.


\section{Selecting Galaxy Properties}\label{Sec:SelectingGalProp}

Galaxies present a variety of physical, spectroscopic and 5-band photometric properties made available in the SDSS-DR8 data catalog. 
We selected the most relevant in order to create a p-Dimensional cloud of properties or features for further study.

Within the photometry-derived properties included are the colors, which show the coarse shape of galaxy spectra, and in some extend the age of the overall stellar population in the galaxy. Only the colors $u-r$ and $g-r$ were selected, since most of the color combinations possible from the $u,g,r,i,z$ bands \citep{fukugita1996} are highly correlated. For computing colors, extinction-corrected model magnitudes \citep{stoughton2002} were used, as well as k-corrections to an observing rest-frame of $z=0$. The k-corrections are calculated by using a template fitting technique used in e.g. \cite{budavari2000} and \cite{csabai2000}. Here, the colors are matched to the colors of a model spectrum defined by a non negative linear combination of redshifted template spectra. Then, the best model spectrum is blueshifted back to the rest frame (z=0) and the k-correction computed. The template spectra are drawn from a list provided by \cite{bruzual2003}.

Since we will study the luminosity function as a function of position in this cloud (Section \ref{Sec:LumfunVsArcLength}), we decided not to include the absolute magnitude $M_{r}$ as a property. If we did, any partitioning of the cloud would introduce non-desired artificial cuts in the range of absolute magnitudes used in the computation of luminosity functions. Therefore, neither the absolute magnitude nor any other strongly correlated property of it (such as stellar mass) should be chosen as part of the properties.

Another photometry-derived feature is the concentration index $C \equiv R90_{r}/R50_{r}$, where $R90_{r}$ and $R50_{r}$ are the radii enclosing the 90 and 50$\%$ of the r-band petrosian flux, respectively. This index has been found to correlate with galaxy morphological type \citep{strateva2001,shimazaku2001}. Indeed, de Vaucouleurs light profiles of elliptical galaxies are more concentrated than the exponential profile in the disks of spiral galaxies. 

The redshift-dependent $r$-band surface brightness defined by $\mu_{50,r} = m_{r} + 2.5\log [2 \pi R50_{r}^{2}(1+z)^4]$ is also included as a property. This breaks the degeneracy of $R90_{r}/R50_{r}$ between bright and dim spiral galaxies. Here we use the extinction and k-corrected petrosian apparent magnitude $m_{r}$, taking $\sqrt{2}R50_{r}$ as a less noisy proxy for the petrosian radius \citep{stoughton2002,strauss2002}.

The physical properties selected are the star formation rate (SFR), specific star formation rate (SFR/$M_{*}$, where $M_{*}$ is the stellar mass) and petrosian r-band mass-to-light ratio ($M_{*}/{\rm L}_{r}$). These are included in SDSS-DR8 and obtained from galaxy spectra analysis at MPA and JHU \footnote{http://www.mpa-garching.mpg.de/SDSS}, as detailed in \cite{kauffmann2003a}, \cite{brinchmann2004}, \cite{tremonti2004}, \cite{gallazzi2005} and \cite{salim2007}.

Note that $M_{*}$ has been derived from template fitting to the total flux in the 5 photometric bands \citep{aihara2011}. As the spectral fibers diameter cover only $3''$ of the central part of each galaxy, the SFR had to be corrected for this deficiency to its full value \citep{brinchmann2004}.  

Since other spectral features such as lick indices or line equivalent widths are non-trivial to extrapolate from their fiber values to the full galaxy ones, we do not include these in the building of the cloud of properties. We do, however, study them apart in Sections \ref{Sec:SpecAndPhysProp} and \ref{Sec:Discussion}.
\\


\section{Methods For Dimensionality Reduction}\label{Sec:MethodDimReduc}

Most of the time, data-mining deals with the data point matrix $\mathbf{A}= [ \mathbf{A}^1 \mathbf{A}^2...\mathbf{A}^p ]$ $\in \mathbb{R}^{N\times p}$, composed by the columns $\lbrace \mathbf{A}^{i} \rbrace_{i=1}^{p}$ that contain $N$ observations for each of the $p$ properties or features. Thus $\mathbf{A}$ can then be thought as a length-$N$ realization of the random vector $\mathbf{X}=[X_{1}...X_{p}] \in \mathbb{R}^{p}$ with distribution $D_{\mathbf{X}}(\mathbf{x})$.

In our work, dimensionality reduction is used for explaining the variations of $\mathbf{X}$ as function of only 1 parameter. For that effect, we use Weighted PCA and Principal Curves, whose detailed descriptions are included in Sections \ref{Sec:PCA} and \ref{Sec:PCurve}.

\subsection{Weighted PCA (WPCA)}\label{Sec:PCA}

Principal Components Analysis (PCA) \citep{pearson1901,jackson1991,jolliffe2002}, also known as Karhunen-Loeve Transform, is a widely used method for dimensionality reduction and classification. It can be seen as a transformation involving a translation, linear scaling and rigid rotation of a collection of $N$ $p$-dimensional data points onto a new coordinate system. The new orthonormal axes, or principal components $\lbrace \mathbf{PC}_{i} \rbrace _{i=1}^{p} \in \mathbb{R}^{N \times 1}$, are constructed such that the projections of the data points on the $\mathbf{PC}_{i}$s are uncorrelated. $\mathbf{PC}_{1}$ is selected as the axis on $\mathbb{R}^{p}$ which has the highest possible variance of the points projected onto it. The next $\mathbf{PC}_{i}$s are ordered in descending value of the variance, having $\mathbf{PC}_{p}$ the lowest. Thus, dimensionality reduction is attained by describing the data in terms of the most important principal components \citep{hastie2009}. This can be obtained by considering only the space spanned by the first $q \leq p $ variance-ranked eigenvectors whose cumulative variance reaches above a high enough threshold.

In practice, the PCs and their variances can be found using singular value decomposition (SVD) of the covariance matrix $\mathbf{C}$ of the data points \citep{golub1996}. SVD allows us to factorize it in the form $\mathbf{C} = \mathbf{V \Sigma V}^{\rm T}$. Here, $\mathbf{\Sigma}$ is a diagonal matrix with the eigenvalues (variances) and $\mathbf{V}$ contains the eigenvectors (principal components) in the respective columns. Thus, $\mathbf{V}$ contains the expansion coefficients of the transformation $\mathbf{PC}_{i} = \Sigma_{j=1}^{j=p} \mathbf{V}_{ji} \mathbf{x}_{j}$ ($i=1,..,p$) from property space to PC space.

In Weighted PCA, the covariance matrix is calculated in a weighted schema. Many times we are confronted with noisy or missing data points. As a solution, we can assign a weight $w_{i}>0$ to each $i$th data point in order to account for the noisy data points or the missing ones. In this context, WPCA involves considering these weights in the calculation of all averages and covariances between the $p$ properties. In general, the properties might have different units, for which they have first to be made unitless by standardization of the data points (subtract to each property its (weighted) average and then divide it by its (weighted) standard deviation).


\subsection{Principal Curves}\label{Sec:PCurve}

Principal curves (P-curves) and surfaces (P-surfaces) \citep{hastie1984,hastie1989,tibshirani1992,gorban2008} go one step ahead of PCA, providing a low-dimensional {\it curved} manifold that passes trough the middle of the data points. In this paper we consider a 1-parameter (called $l$) principal curve $\mathbf{f}(l)=[f_{1}(l),...,f_{p}(l)] \in \mathbb{R}^{N \times 1}$, where each of the $N$ data points $\mathbf{x}=[x_{1}...x_{p}]$ is given a unique closest projection $\mathbf{f}(l_{\mathbf{f}}(\mathbf{x}))$ onto the curve. As a convention, $l_{\mathbf{f}}(\mathbf{x})$ is chosen to be the arc length from the beginning of the curve to the projection point of $\mathbf{x}$. Under this context, the P-curve can be considered by itself the 1st and only $curved$ principal component, as the dimensionality of the data is reduced from $p$ to 1 dimension. In practice, the P-curve is composed by $N-1$ line segments that connect the projection points.

The principal curve is defined as the {\it average} of the data-points that project onto it, minimizing the projection distance between $\mathbf{x}$ and $\mathbf{f}(l_{\mathbf{f}}(\mathbf{x}))$ over all points. This property of self consistency allows us to follow a series of iterative projection-expectation steps for its construction \citep{hastie1989}. In fact, an educated first guess for the P-curve is to make it equal to $\mathbf{PC}_{1}$. Later on, the $j$th estimate ${f}^{(j)}_{i}(l)$ of the curve at the $j$th expectation step is calculated as ${f}^{(j)}_{i}(l) = E(\mathbf{X}_{i}\vert {l}_{\mathbf{f}^{(j-1)}}(\mathbf{X})= l) $. In practice, we compute this expression using a weighted penalized cubic B-spline regression \citep{silverman1985,hastie1990,ruppert2003,hastie2009}. These splines are calculated on a series of $k$ knots chosen from the data points, while the degrees of freedom ($df$) of the regression control the degree of smoothing of the P-curve. On the other hand, the $j$th projection step is performed next, involving the search for the closest perpendicular projection of $\mathbf{x}$ onto $\mathbf{f}^{(j)}(l)$, which is composed by the $N-1$ line segments. The iterations stop when the cumulative projection distances from the data points to the P-curve do not change significantly with respect to the one in the previous step.

Although P-curves are constructed on the $p$-dimensional space of properties, we can consider building the P-curve of the data points projected on the first $q$ most important principal components of the WPCA. This minimizes the complexity and computations, specially in the case $p\gg q$, without losing much information. The approximation is of course valid as long as the first $q$ eigenmodes contain as much of the total variance as possible.


\section{Building the Principal Curve and Population Separators along Arc Length}\label{Sec:BuildingPcurve}


\subsection{$V_{\rm max}$ weighting}\label{Sec:Vmax}

As the MGS is a magnitude limited sample, not all galaxy types are sampled equally in the survey volume. As a consequence, we used WPCA and a weighted principal curve of the galaxy population to get an unbiased result.

In detail, at higher redshifts we sample mostly the brightest galaxies, neglecting the faint ones (Malmquist bias). On the other side, at low redshifts the SDSS spectrograph fails to take the spectra of very bright galaxies (see Section \ref{Sec:GalaxySample}). 

As a solution, we use the $V_{\rm max}$ weighting method \citep{schmidt1968} to account for this incompleteness. Here, each $i$-th galaxy is assigned a weight $w_{i}=V_{\rm S}/V_{{\rm max},i}\geq 1$, where $V_{\rm S}$ is the volume of the survey. Here we note that, given the particular $[z_{1},z_{2}]$ and $[m_{1},m_{2}]$ intervals for the survey, the $i$-th galaxy found at $z_{i}$ could be observed only within a maximum comoving volume $V_{{\rm max},i} \leq V_{\rm S}$. If the $i$-th galaxy of apparent magnitude $m_{i}$,  k-correction $k_{i}=k(z_{i})$, and at a luminosity distance $D_{L}(z_{i})$ were to have limiting apparent magnitudes $m_{1,2}$, then it should be moved to a limiting luminosity distance $D_{L,i}(m_{1,2})$ given by 

\begin{eqnarray}
D_{L,i}(m_{1,2}) \equiv D_{L,i}(z_{\rm lim};m_{1,2}) & \nonumber \\  
=  D_{L}(z_{i}) \times  10^{(m_{1,2} - k(z_{\rm lim}) - m_{i} + k_{i})/5}. &  \label{Eq:DlumLim}
\end{eqnarray}
Hence, the maximum volume is defined by the biggest interval of $D_{L}$ inside which a galaxy can appear in the survey:
\begin{eqnarray}
V_{{\rm max},i} & = & [V(\min (D_{L}(z_{2}),D_{L,i}(z_{\rm lim};m_{2})   )) \nonumber \\
					 & - & V(\max (D_{L}(z_{1}),D_{L,i}(z_{\rm lim};m_{1})   ))] \times F_{A},
\label{Eq:Vmax}
\end{eqnarray}
As Eq. \eqref{Eq:DlumLim} defines $z_{\rm lim}$ in an implicit way, we solve for it iteratively. We calculated the $V_{\rm max}$ values directly inside the database using an integrated cosmological functions library \citep{taghizadeh-popp2010}. 

The PCA, P-curve and calculations related to volume densities in this paper (such as histograms) are all $V_{\rm max}$ weighted.


\subsection{WPCA results}\label{Sec:WPCA}

As measure for not skewing the PCA, we clipped off visually the outliers in each of the $p=7$ galaxy properties in order to dismiss artifacts or wrong measurements. We also used only galaxies which have all the 7 properties measured, simplifying the calculations and avoiding using Gappy PCA \citep{connolly1999}. This left us with a total of $N=171,698$ galaxies (99\% of the initial ones).

\begin{figure}[ht]
\epsscale{1.0}
\plotone{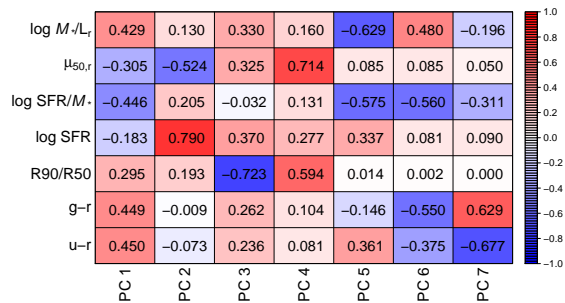}
\caption{The $\mathbf{V}$ matrix resulting from applying WPCA on the 7 galaxy properties. The columns are the orthonormal principal components (i.e. eigenvectors of the covariance matrix). Each $\mathbf{PC}_{i}$, $i=1,...,7$ can be viewed as a linear combination of properties, with the expansion coefficients $\mathbf{V}_{ji}$ of the $j$th property stored in the $j$th row. Coefficients with stronger color show a higher importance of the property for the given PC. The sign of the coefficient shows correlations/anticorrelations between the properties and the PC.}
\label{Fig:PCAEigenvectors}
\end{figure}

\begin{table}[h]
\begin{center}
\caption{WPCA variances}
\begin{tabular}{cccccccc}\tableline 
 & $\mathbf{PC}_{1}$ & $\mathbf{PC}_{2}$ & $\mathbf{PC}_{3}$ & $\mathbf{PC}_{4}$ & $\mathbf{PC}_{5}$ & $\mathbf{PC}_{6}$ & $\mathbf{PC}_{7}$\\ 
\tableline
$\sigma_{\mathbf{PC}}^2$ &  4.493 &  1.115 &  0.842 &  0.363 &  0.090 &  0.068 &  0.030\\ 
$\sum \frac{\sigma_{\mathbf{PC}}^2}{p}$ &  0.642 &  0.801 &  0.921 &  0.973 &  0.986 &  0.996 &  1.000\\ 
\tableline
\label{Table:PCAeigenvalues}
\end{tabular}
\end{center}
{\small Variance for each principal components and its associated cumulative variance. Since the data has been standardized, the sum of the variances is equal to the number of dimensions ($p=7$). }
\end{table}

\begin{figure*}[ht!]
\epsscale{1.0}
\plotone{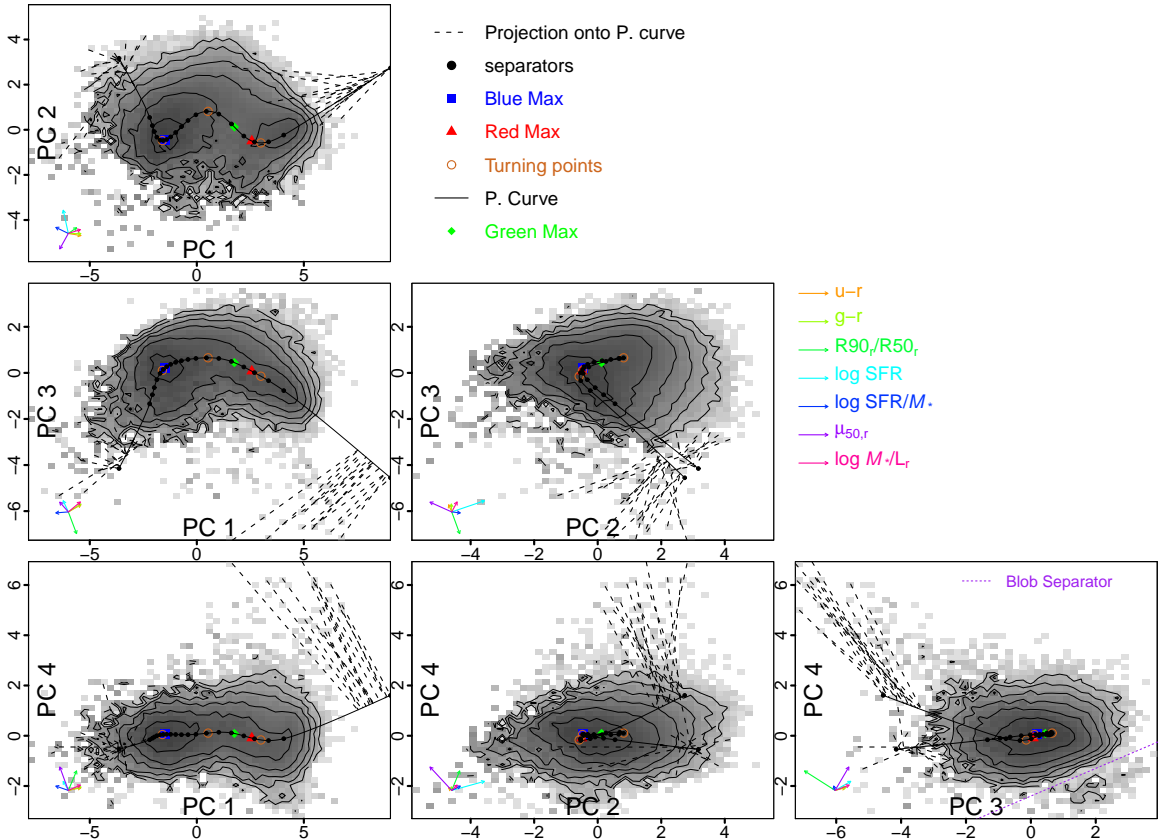}
\caption{Principal curve (black continuous line) fitted to the first 4 principal components (density maps are log-scaled, with contour curves separated by 0.5 dex). The arc length increases in the direction of increasing $\mathbf{PC}_{1}$. The first and last 15 data points (ordered by arc length) are connected to their corresponding projections on the curve with dashed lines. The separators between the $\lbrace L_{i}\rbrace_{i=1}^{i=20}$ groups are shown as black circles on top of the curve (see text). The curve presents 3 turning points marked as brown rings. The colored arrows show the directions and relative strength of the galaxy properties projected onto PC space. In $\mathbf{PC}_{3}$ v/s $\mathbf{PC}_{4}$, we plotted the separating line between the main cloud and a small blob of galaxies, given by  $\mathbf{PC}_{4} \leq -1.3 + 0.55(\mathbf{PC}_{3}-2.0)$ (see Sec. \ref{Sec:RedSpiralsPC3PC4}).}
\label{Fig:Pcurve4DturningPoints} 
\end{figure*}

Figure \ref{Fig:PCAEigenvectors} and Table \ref{Table:PCAeigenvalues} present the results from computing WPCA on the 7 galaxy properties.
From Table \ref{Table:PCAeigenvalues}, we can notice that most of the information  (97\% of the total variance) is contained in the first 4 principal components. On Fig. \ref{Fig:PCAEigenvectors}, each $\mathbf{PC}_{i}$, $i=1,...,7$ can be viewed as a linear combination of properties, with the expansion coefficients $\mathbf{V}_{ji}$ of the $j$th property stored in the $j$th row. Coefficients with stronger color show a higher importance of the property for the given PC. The sign of the coefficient shows correlations/anticorrelations between the properties and the final value of the PC.

For $\mathbf{PC}_{1}$, the strength (absolute magnitude) of its expansion coefficients $\mathbf{V}_{j 1}$ in the basis of the galaxy properties is shared mostly evenly between these properties, being $u-r$, $g-r$, $SFR/M_{*}$ and $M_{*}/L_{r}$ the most important. The correlations show that high values of $u-r$, $g-r$, $M_{*}/L_{r}$ and $R90_{r}/R50_{r}$, together with low values of $\mu_{50,r}$, $SFR/M_{*}$ and $SFR$, will produce a high $\mathbf{PC}_{1}$ value. We might therefore expect that $\mathbf{PC}_{1}$ is a good separator between the young, blue population of spirals/irregulars and the old population of red old ellipticals. 

For $\mathbf{PC}_{2}$, the $SFR$ and $\mu_{50,r}$ are the most important, having opposite signs. Thus, we expect galaxies with bright surface brightness and high star formation to show high values of $\mathbf{PC}_{2}$.

For $\mathbf{PC}_{3}$, the most important property is $R90_{r}/R50_{r}$, with an opposite correlation with respect to the next important properties of mostly equal strength ($u-r$, $g-r$, $SFR$, $M_{*}/L_{r}$ and $\mu_{50,r}$). We can expect that big and bright star-forming spiral galaxies with reddish colors (probably from a red core) should have high $\mathbf{PC}_{3}$.

For $\mathbf{PC}_{4}$, all the properties have the same correlations, being $\mu_{50,r}$, $R90_{r}/R50_{r}$ and $SFR$ the most important. Thus, concentrated (and possibly star-forming) galaxies of faint surface brightness have high values of $\mathbf{PC}_{4}$. As the variance along $\mathbf{PC}_{4}$ is much smaller than the previous PCs, this is a rare combination of correlation for these properties to be observed at the same time.

Furthermore, the last 3 PCs ($\mathbf{PC}_{5}$,$\mathbf{PC}_{6}$ and $\mathbf{PC}_{7}$), which account for less than 2\% of the total variance, are less obvious to interpret. They might trace either special cases of galaxy populations, or just artifacts and wrong/noisy measurements of the properties.


\begin{figure*}
\epsscale{1.00}
\plotone{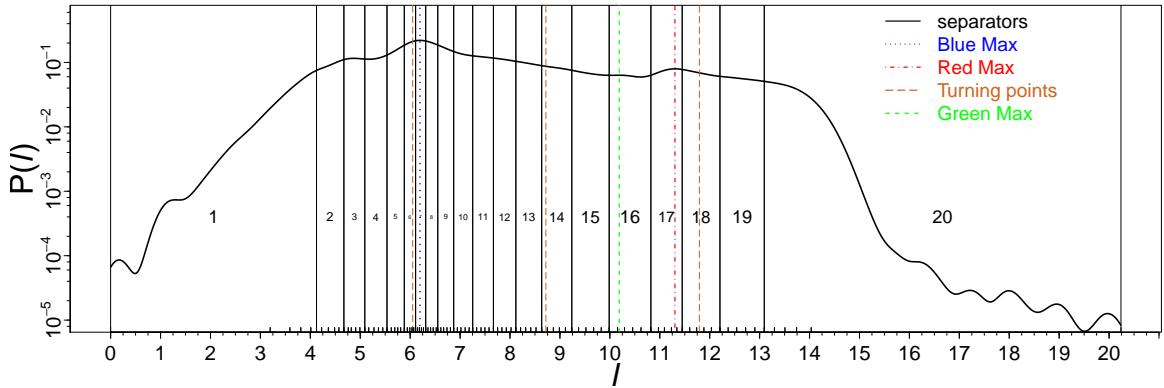}
\caption{Probability density of the arc-length values $\lbrace l_{i}\rbrace _{i=1}^{i=N}$ measured at the points of the curve where the $N$ data points are projected onto. The arc-length increases in the direction of increasing $\mathbf{PC}_{1}$. Vertical black continuous lines denote the population separators, while the numbers denote the $\lbrace L_{i}\rbrace_{i=1}^{i=20}$ galaxy groups. The 5 small tick marks within each galaxy group mark the boundaries of the subgroups $\lbrace \lambda_{1},...,\lambda_{100}\rbrace$.}
\label{Fig:LambdaDistribution}
\end{figure*}

\subsection{The fitted Principal Curve and Population Separators along it}\label{Sec:PrincipalCurveAndPopSep}

\begin{figure}[t]
\epsscale{1.00}
\plotone{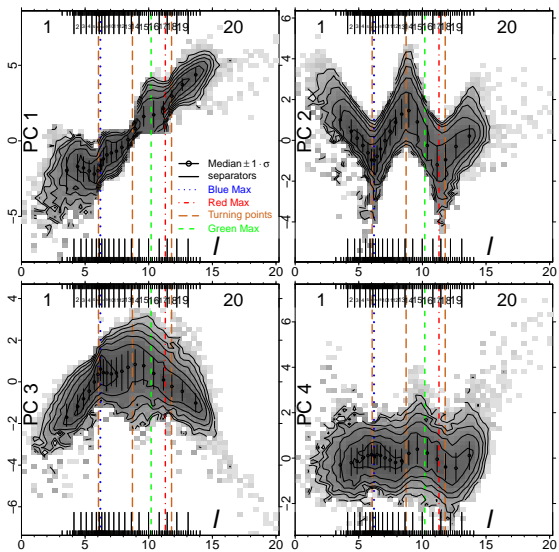}
\caption{ Density maps of the principal components (y-axis) as a function of the arc-length $l$ (x-axis). Density is log-scaled, contour curves separated by 0.5 dex. Population separators are shown as vertical tick marks. The numbers denote the $\lbrace L_{i}\rbrace_{i=1}^{i=20}$ galaxy groups. Colored vertical lines show the position of the maxima and turning points at particular $l$ values.}
\label{Fig:PCAvsLambda} 
\end{figure}

We decided to construct the Principal curve in the 4-dimensional space defined by $\lbrace\mathbf{PC}_{1},...,\mathbf{PC}_{4}\rbrace$, since their combined cumulative variance (0.973) is close to unity (see Table \ref{Table:PCAeigenvalues}). Although the computation for the number of dimensions and data points involved is not too intensive, we think of this as a pedagogical example that can be used for other extreme cases when $N\geq 10^{10}$ objects with $p\geq 100$ dimensions, for instance. In fact, our election does not change significantly the results compared to using $p=7$.

In the expectation step for creating the principal curve, each $\mathbf{PC}_{i}$ is fitted with penalized B-splines of 5.4 degrees of freedom ($df$), defined at a sequence of $k=211$ unique knots chosen at equally spaced quantiles of arc-length values.
Principal curves with $df \gtrsim 7$ make the curve to oscillate excessively, turning back and forth across and along $\mathbf{PC}_{1}$, whereas with $df \simeq 4$ resemble more closely a straight line along the $\mathbf{PC}_{1}$ direction.

Figure \ref{Fig:Pcurve4DturningPoints} shows the result of fitting the principal curve to $\lbrace\mathbf{PC}_{1},...,\mathbf{PC}_{4}\rbrace$. The 4-dimensional cloud of properties presents 2 density maxima placed mainly along the $\mathbf{PC}_{1}$ direction, corresponding to the blue and red galaxy populations. The principal curve mostly resembles the letter "W", presenting clearly 4 different regimes or branches separated by 3 turning points (T-points).

We created 20 equal number density galaxy groups (in Mpc$^{-3}$) labeled as $\lbrace L_{i}\rbrace_{i=1}^{i=20}$ by placing population separators at fixed arc length values along the P-curve, as shown in Figures \ref{Fig:Pcurve4DturningPoints} and \ref{Fig:LambdaDistribution}. Galaxies are grouped together into the same $L_{i}$ group when the arc length values measured at their projections points onto the P-curve are placed between 2 consecutive separators. These separators are positioned in such a way that the ($V_{\rm max}$-weighted) number density (in Mpc$^{-3}$) of the galaxies belonging to each of the 20 $L_{i}$ groups amounts to 1/20th of that from the whole sample of galaxies. This allowed us to study the 4 principal curve branches in detail. We chose the arc-length to increase in the same direction of increasing $\mathbf{PC}_{1}$, with growing values of arc length as we progress from $L_{1}$ to $L_{20}$. Thus, the P-curve's 1st branch is comprehended in $\lbrace L_{1},...,L_{6} \rbrace$, the 2nd branch in $\lbrace L_{7},...,L_{14} \rbrace$, the 3rd branch in $\lbrace L_{15},...,L_{18} \rbrace$ and the 4th branch in $\lbrace L_{19},L_{20} \rbrace$. Table \ref{Table:20GroupStatistics2} shows some statistics of these groups. 

Within each $L_{i}$ group, we further created 5 subgroups of galaxies along arc length naming them $\lbrace \lambda_{i} \rbrace_{i=1}^{i=100}$, also of equal number density in Mpc$^{-3}$ as explained before. We further partitioned these groups similarly, now using several radial separators in the perpendicular direction to the curve, defining 10 concentric cylinder-like separating surfaces. In this way, the groups defined by this finer partitioning have all the same number density (in Mpc$^{-3}$), equal approximately to 1/1000th of the number density of the whole sample. This allowed us to identify and extract localized galaxy populations positioned very close to the spine of the cloud of properties, and study them in Sec. \ref{Sec:RedSpineGaussianLumFun}.

Figure \ref{Fig:LambdaDistribution} shows the probability density distribution of the arc-length $l$ values, as well as the population separators. The curve has a length of $l_{max}$=20.24, and the variance of the arc length values is $\sigma_{l}^2=7.79$, measured with respect to the center of the curve at $<l>=7.91$. Note that Table \ref{Table:20GroupStatistics2} shows that the quadratic mean (root mean square) of all the projection distances from the data points to the P-curve takes a value of $d_{\perp}=1.31$, which is small compared to the length of the curve. The blue and red peaks of maximum density are clearly visible, as well as a small green peak. The 1st turning point (at $L_{6}$) lies closely with the blue maximum ($L_{7}$), whereas the red maximum ($L_{17}$) is a little behind of the 3rd T-point ($L_{18}$), after which we can find a hump defining the red sequence of galaxies. We find a green maximum ($L_{16}$) standing in between the 2nd T-point ($L_{14}$) and the red maximum.

Figure \ref{Fig:PCAvsLambda} shows the density maps of the scatter of each $\lbrace\mathbf{PC}_{1},...,\mathbf{PC}_{4}\rbrace$ as a function of the arc-length.  
The different shapes that this scatter presents depend evidently on the contortions or twists of the principal curve along the PCs. 
As the 4 branches of the curve mostly turn left and right along  $\mathbf{PC}_{2}$, the scatter in $\mathbf{PC}_{2}$ show the same "W" shape as the P-curve. On the other hand, the curve increases its length into the $\mathbf{PC}_{1}$ direction, so the scatter shows a mostly linear relation between $\mathbf{PC}_{1}$ and arc length. The same analysis applies to the scatter of the next PCs, which is boomerang-shaped for $\mathbf{PC}_{3}$ and mostly constant with respect to arc-length for $\mathbf{PC}_{4}$ (although with little wiggles).

\begin{table}[ht!]
\begin{center}
\caption{Statistics of the $\lbrace L_{i} \rbrace _{i=1}^{i=20}$ galaxy group.\rlap{$^a$}}
\begin{tabular}{lrrrrr}\tableline 
\tableline
Group & $N_{gal}$ & $l_{\rm min}$ & $l_{\rm max}$ & $<l>$ & $d_{\perp}$\\ 
\tableline
$L_{1}$ & 4050 & 0 & 4.12 & 3.54 & 1.57\\ 
$L_{2}$ & 2987 & 4.12 & 4.67 & 4.41 & 1.34\\ 
$L_{3}$ & 2368 & 4.67 & 5.09 & 4.87 & 1.25\\ 
$L_{4}$ & 2136 & 5.09 & 5.54 & 5.32 & 1.14\\ 
$L_{5}$ & 1589 & 5.54 & 5.88 & 5.72 & 1.12\\ 
$L_{6}$ & 1345 & 5.88 & 6.11 & 6.01 & 1.17\\ 
$L_{7}$ & 1674 & 6.11 & 6.31 & 6.21 & 1.27\\ 
$L_{8}$ & 2190 & 6.31 & 6.55 & 6.44 & 1.29\\ 
$L_{9}$ & 3568 & 6.55 & 6.87 & 6.71 & 1.12\\ 
$L_{10}$ & 6287 & 6.87 & 7.25 & 7.06 & 1.16\\ 
$L_{11}$ & 10196 & 7.25 & 7.67 & 7.46 & 1.21\\
$L_{12}$ & 13862 & 7.67 & 8.12 & 7.89 & 1.31\\ 
$L_{13}$ & 17062 & 8.12 & 8.63 & 8.37 & 1.39\\ 
$L_{14}$ & 18283 & 8.63 & 9.24 & 8.93 & 1.51\\ 
$L_{15}$ & 15287 & 9.24 & 9.99 & 9.6 & 1.56\\ 
$L_{16}$ & 9421 & 9.99 & 10.82 & 10.39 & 1.50\\ 
$L_{17}$ & 5546 & 10.82 & 11.45 & 11.16 & 1.51\\ 
$L_{18}$ & 9610 & 11.45 & 12.21 & 11.82 & 1.34\\ 
$L_{19}$ & 19877 & 12.21 & 13.09 & 12.64 & 1.12\\ 
$L_{20}$ & 24360 & 13.09 & 20.24 & 13.69 & 1.11\\ 
\tableline
All & 171698 & 0 & 20.24 & 7.91 & 1.31 \\ 
\tableline
\label{Table:20GroupStatistics2}
\end{tabular}
\end{center}
{\small $^a$~ $N_{gal}$ denotes the number of galaxies in each group, comprehended in the arc length interval  $[l_{\rm min},l_{\rm max}]$ of $<l>$ average arc length. The value $d_{\perp}$ denotes the quadratic mean (root mean square) of the projection distances from the data points onto the P-curve.}
\end{table}


\section{Galaxy properties and statistics as function of Arc Length}\label{Sec:ArcLengthDependance}

In this section we show how galaxy properties, luminosity functions and spatial clustering change as a function of the $\lbrace L_{i}\rbrace_{i=1}^{i=20}$ equal number density galaxy groups (ordered in ascending arc-length).

Compared to $\mathbf{PC}_{1}$ alone, the principal curve provides much more information about particular changes in properties along its arc length. We will see that the evolution of galaxy properties along the curve is intimately related to the "W" shape of the principal curve, where each of the 4 branches define particular galaxy populations.


\begin{figure*}
\begin{center}
\epsscale{1.8}
\plottwo{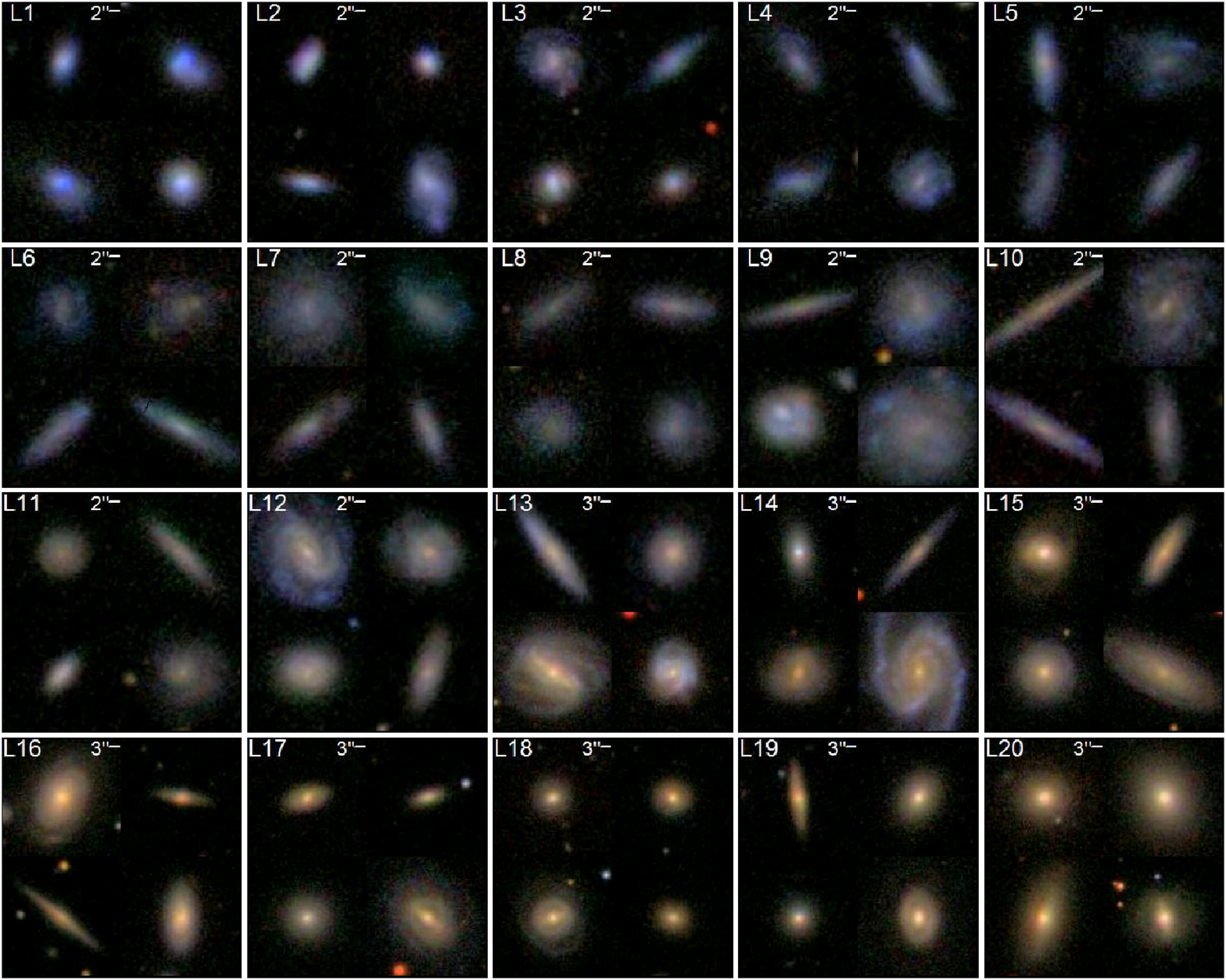}{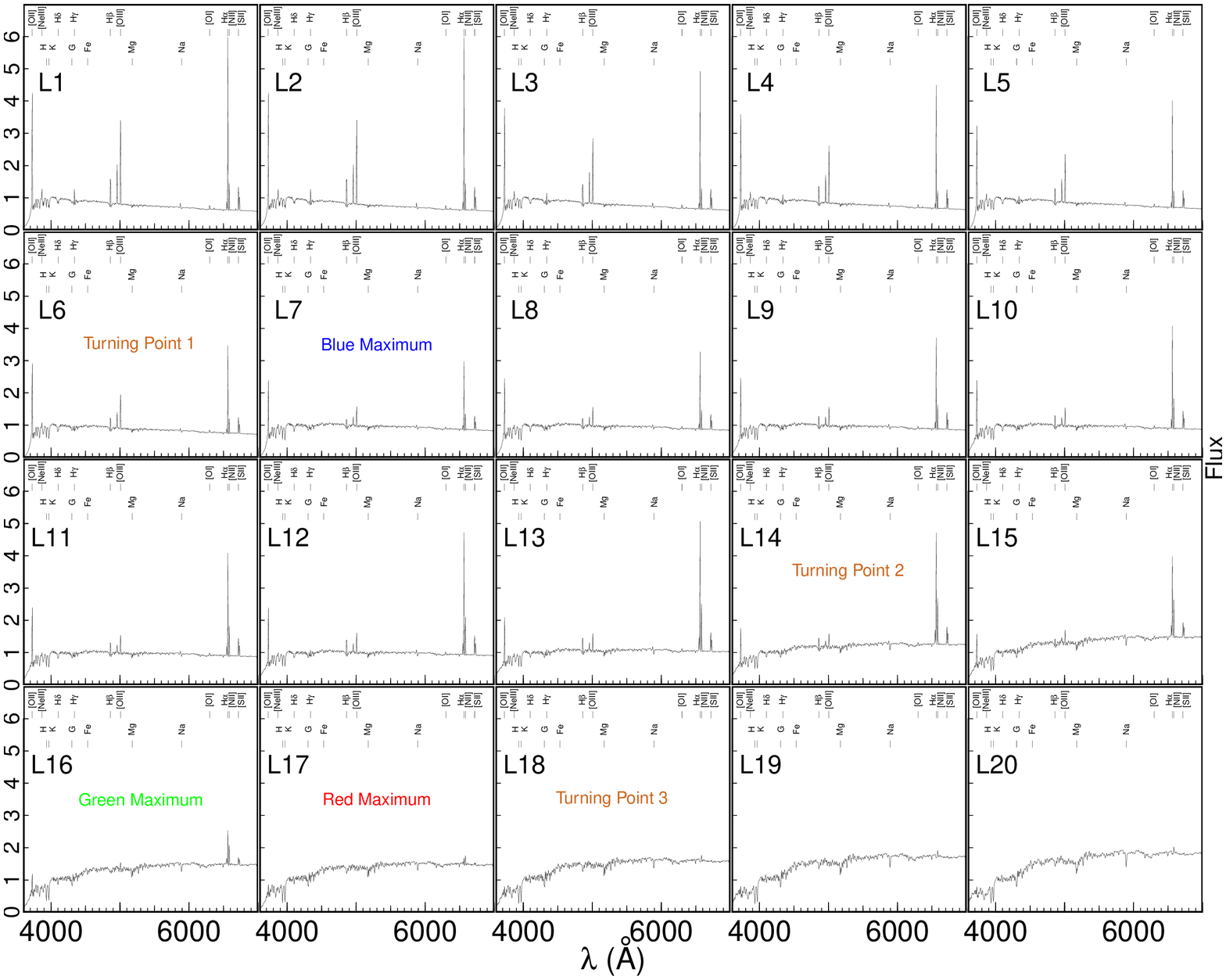}
\caption{\emph{Top}:Panels with the 4 most representative galaxy shapes that appear in each of the $\lbrace L_{i}\rbrace_{i=1}^{i=20}$ galaxy groups. Arc-length increases from $L_{1}$ to $L_{20}$  White bars show the scale in arcseconds.  \emph{Bottom}: Average rest-frame spectrum of same 20 groups as above. The flux (y-axis) is normalized to be 1 at $\lambda=4000$ \AA. The average of each group was performed on 1000 galaxies sampled randomly with probabilities proportional to $V_{\rm max}^{-1}$. Also marked are the positions of emission lines (top row) and absorption lines (bottom row).   }
\label{Fig:MorphologyAndGalSpectra} 
\end{center}
\end{figure*}

\subsection{Morphology and Average Spectra}\label{Sec:MorphologyAndGalSpectra}

Figure \ref{Fig:MorphologyAndGalSpectra} shows the most representative galaxy morphologies and average spectra for the $\lbrace L_{i}\rbrace_{i=1}^{i=20}$ groups.

The most evident feature is the change in color and the slope of the spectra (from blue to red), as well as an overall weakening of emission lines (e.g. Balmer series of Hydrogen and forbidden lines, such as OIII, OII, NII, etc.) and an increase of metallic absorption lines and bands (Na,Mg,H,K,G) as we reach high arc length values. In the same way, morphological types include various types of blue galaxies at the beginning and middle of the curve, whereas red ellipticals dominate the end of it. This bimodality is expected and agrees with $\mathbf{PC}_{1}$ in Fig. \ref{Fig:PCAEigenvectors}, appearing also other studies as the change along the 1st principal component \citep[e.g][]{yip2004b,coppa2011}. We can, however, identify as well more subtle populations along arc length, not distinguishable in $\mathbf{PC}_{1}$ alone. These distinct population are defined on each of the 4 branches of the principal curve, connected by the 3 turning points.

With respect to morphologies, we see that the arc length correlates very well with the Hubble galaxy type. We however miss the distinction between barred/non-barred spiral galaxies due to the lack of properties able to separate them. Blue irregulars and blue compact dwarf (BCD) galaxies \citep{papaderos2006,corbin2006} appear in the 1st branch of the principal curve. Some of these type of BCDs were identified as the green pea galaxies at higher redshift \citep{cardamone2009}. These morphologies change then into low surface brightness galaxies (LSBGs) with spiral and irregular shapes, which dominate the 1st turning point and blue maximum. Bright spirals with strong blue star forming arms appear in the second branch, which by the 2nd turning point show sizable bulges. A dramatic change happens in the 3rd branch, where reddish big-bulged spirals and lenticulars dominate, forming part of the green and red maxima. A new transition happens at the 3rd turning point, having the big bright red ellipticals (CDs) and brightest cluster galaxies (BCGs) dominate at the end of the P-curve's 4th branch.

Emission lines, such as the forbidden OII, OIII, SII and NeII, as well as the Balmer series of Hydrogen (e.g. H$_{\alpha}$, H$_{\beta}$, H$_{\gamma}$), are strong in the violently starforming blue galaxies at the 1st branch. These lines weaken as we transition into LSBGs, but interestingly H$_{\alpha}$ and H$_{\beta}$ become stronger in the 2nd branch, reaching maximal values in the starforming spirals at the 2nd turning point. After this, they weaken again to become imperceptible in the bright ellipticals in the 4th branch. NII follows the same  pattern as H$_{\alpha}$, but somehow remains still visible in CD galaxies, as seen in many spectral atlas \citep[e.g.][]{dobos2012}. On the other hand, OIII declines steadily through arc length, disappearing after the red maximum.

Absorption lines, such as Na, Mg and the G band become evident in the starforming spirals by the end of the 2nd branch (as the bulge increases in size), and appear strong in the ellipticals at the 4th branch. Although the H and K lines of calcium are always visible, the 4000\AA break increases steadily with arc length, turning into a striking feature in bright ellipticals.


\begin{figure*}
\plotone{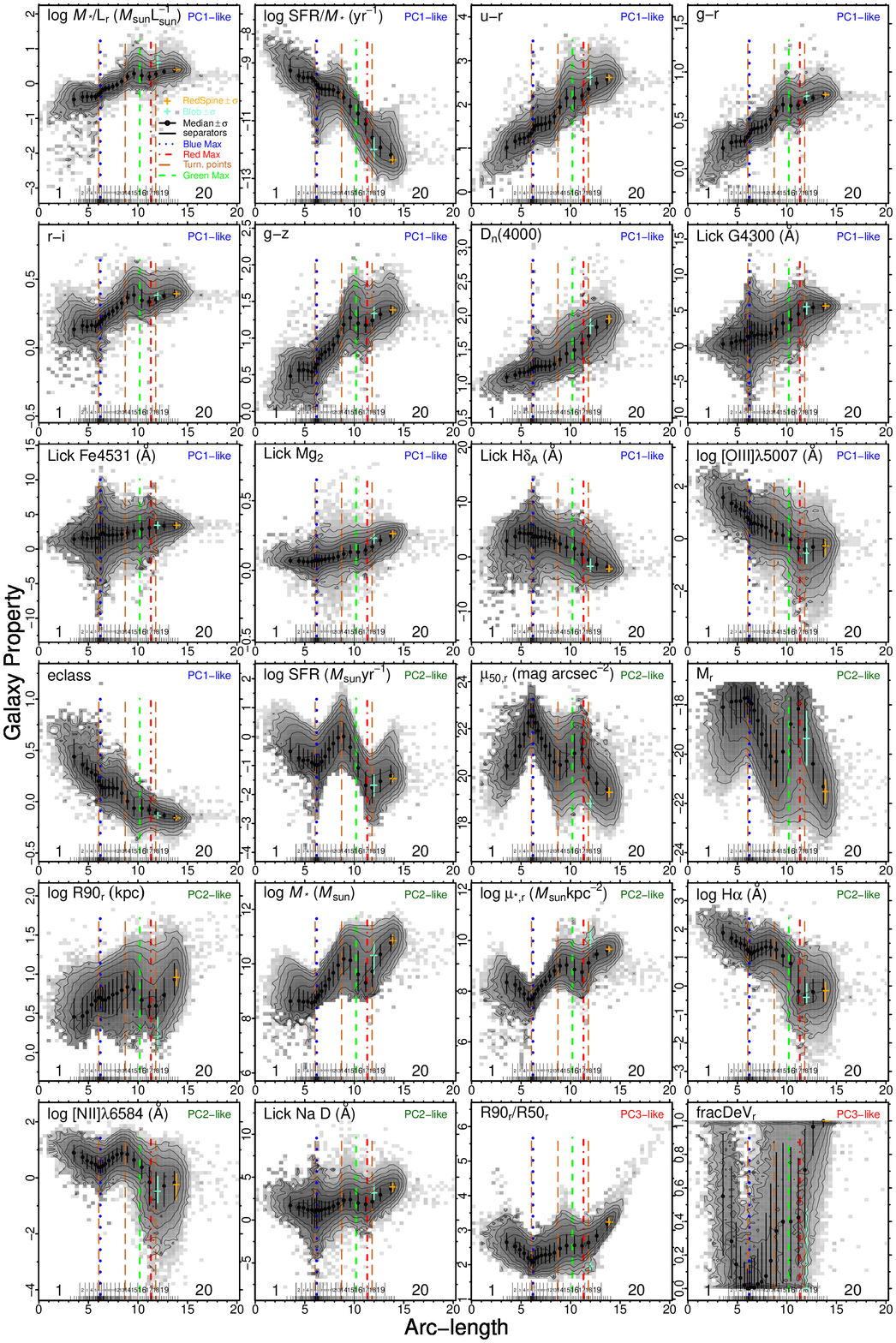}
\caption{Galaxy properties (y-axis) as a function of arc-length in the principal curve (x-axis). Properties are ordered row-wise, and grouped with respect to which principal component they look like the most as in Fig. \ref{Fig:PCAvsLambda} (upper right hand corner of each panel). The $\mathbf{PC}_{1}$ case resembles a straight line, $\mathbf{PC}_{2}$ a "W" and $\mathbf{PC}_{3}$ a boomerang. The black circles at the mean arc length value within each $\lbrace L_{i} \rbrace_{i=1}^{i=20}$ group show the position of the median of the distribution of the property in it, together with vertical bars spanning the 15.9\% to 84.1\% quantiles ($\pm 1 \sigma$). The orange and cyan bars show respectively the same quantiles for the red spine and red spiral blob galaxies discussed in Sec. \ref{Sec:InterestingGroups}.}
\label{Fig:LambdaVsProp}
\end{figure*}

\begin{table*}[ht!]
\renewcommand{\arraystretch}{1.3}
\begin{center}
\caption{Medians of galaxy property distributions in each galaxy group, together with the 15.9\% to 84.1\% quantiles ($\pm 1 \sigma$) .}
\scriptsize \addtolength{\tabcolsep}{-4pt}
\begin{tabular}{lrrrrrrrrrrrr}
\tableline
\tableline
{\tiny \rm Group} & $\log M_{*}/{\rm L}_{r}$ & log SFR/$M_{*}$ & $u-r$ & $g-r$ & $r-i$ & $g-z$ & $D_{n}(4000)$ & Lick G4300 & Lick Fe4531 & Lick Mg$_2$ & Lick Na D & Lick H$\delta_{\rm A}$ \\ 
	   & ($M_{\odot}L^{-1}_{\odot,r}$) & (yr$^{-1}$)  &       &       &       &       &               &  (\AA)       &    (\AA)      &             &   (\AA)     &      (\AA) \\ 
\tableline
\tableline
$L_{1}$ & $-0.46^{+0.19}_{-0.21}$  & $-9.26^{+0.41}_{-0.24}$  & $1.01^{+0.17}_{-0.24}$  & $0.21^{+0.09}_{-0.11}$  & $0.13^{+0.07}_{-0.06}$  & $0.48^{+0.14}_{-0.18}$  & $1.10^{+0.06}_{-0.08}$  & $0.25^{+0.78}_{-0.70}$  & $1.42^{+0.82}_{-0.87}$  & $0.07^{+0.02}_{-0.02}$  & $1.72^{+1.04}_{-0.78}$  & $2.91^{+1.95}_{-2.96}$ \\ 
$L_{2}$ & $-0.38^{+0.16}_{-0.15}$  & $-9.45^{+0.25}_{-0.20}$  & $1.16^{+0.15}_{-0.17}$  & $0.27^{+0.08}_{-0.09}$  & $0.16^{+0.05}_{-0.06}$  & $0.56^{+0.12}_{-0.16}$  & $1.13^{+0.06}_{-0.06}$  & $0.39^{+1.04}_{-1.17}$  & $1.62^{+0.98}_{-1.07}$  & $0.07^{+0.03}_{-0.02}$  & $1.48^{+0.65}_{-0.95}$  & $3.72^{+1.49}_{-2.37}$ \\ 
$L_{3}$ & $-0.38^{+0.12}_{-0.14}$  & $-9.51^{+0.24}_{-0.15}$  & $1.21^{+0.12}_{-0.15}$  & $0.28^{+0.07}_{-0.07}$  & $0.15^{+0.05}_{-0.05}$  & $0.57^{+0.11}_{-0.15}$  & $1.16^{+0.06}_{-0.07}$  & $0.63^{+1.18}_{-1.19}$  & $1.48^{+0.94}_{-1.32}$  & $0.06^{+0.03}_{-0.02}$  & $1.41^{+0.91}_{-0.99}$  & $4.35^{+1.45}_{-1.58}$ \\ 
$L_{4}$ & $-0.38^{+0.10}_{-0.12}$  & $-9.50^{+0.18}_{-0.18}$  & $1.23^{+0.13}_{-0.15}$  & $0.28^{+0.06}_{-0.06}$  & $0.16^{+0.04}_{-0.04}$  & $0.55^{+0.11}_{-0.13}$  & $1.17^{+0.07}_{-0.06}$  & $0.72^{+1.28}_{-1.41}$  & $1.53^{+1.53}_{-1.46}$  & $0.06^{+0.03}_{-0.03}$  & $1.21^{+1.10}_{-1.00}$  & $4.25^{+1.51}_{-1.76}$ \\ 
$L_{5}$ & $-0.38^{+0.08}_{-0.09}$  & $-9.51^{+0.15}_{-0.17}$  & $1.27^{+0.09}_{-0.14}$  & $0.28^{+0.05}_{-0.05}$  & $0.16^{+0.04}_{-0.04}$  & $0.54^{+0.12}_{-0.16}$  & $1.18^{+0.08}_{-0.07}$  & $0.72^{+1.89}_{-1.98}$  & $1.60^{+1.99}_{-2.23}$  & $0.06^{+0.03}_{-0.04}$  & $1.18^{+1.42}_{-1.61}$  & $4.25^{+1.92}_{-2.02}$ \\ 
$L_{6}$ & $-0.32^{+0.06}_{-0.07}$  & $-9.62^{+0.12}_{-0.17}$  & $1.34^{+0.09}_{-0.10}$  & $0.32^{+0.03}_{-0.04}$  & $0.17^{+0.03}_{-0.04}$  & $0.59^{+0.09}_{-0.11}$  & $1.21^{+0.10}_{-0.08}$  & $1.30^{+2.33}_{-2.41}$  & $1.58^{+2.13}_{-2.27}$  & $0.07^{+0.05}_{-0.05}$  & $0.92^{+1.54}_{-1.84}$  & $4.23^{+2.09}_{-2.41}$ \\ 
$L_{7}$ & $-0.26^{+0.06}_{-0.05}$  & $-9.76^{+0.13}_{-0.26}$  & $1.43^{+0.14}_{-0.09}$  & $0.36^{+0.03}_{-0.04}$  & $0.19^{+0.03}_{-0.04}$  & $0.64^{+0.10}_{-0.10}$  & $1.23^{+0.11}_{-0.08}$  & $1.33^{+2.43}_{-2.88}$  & $2.13^{+2.68}_{-2.70}$  & $0.07^{+0.04}_{-0.05}$  & $1.21^{+1.68}_{-1.83}$  & $3.61^{+2.15}_{-2.16}$ \\ 
$L_{8}$ & $-0.19^{+0.08}_{-0.05}$  & $-9.87^{+0.16}_{-0.43}$  & $1.51^{+0.16}_{-0.12}$  & $0.40^{+0.07}_{-0.04}$  & $0.20^{+0.03}_{-0.04}$  & $0.72^{+0.10}_{-0.08}$  & $1.25^{+0.12}_{-0.10}$  & $1.83^{+1.60}_{-2.26}$  & $2.33^{+2.11}_{-2.44}$  & $0.08^{+0.04}_{-0.04}$  & $1.08^{+1.17}_{-1.22}$  & $3.81^{+1.88}_{-3.04}$ \\ 
$L_{9}$ & $-0.15^{+0.07}_{-0.07}$  & $-9.90^{+0.17}_{-0.22}$  & $1.53^{+0.18}_{-0.13}$  & $0.41^{+0.06}_{-0.05}$  & $0.23^{+0.03}_{-0.03}$  & $0.78^{+0.08}_{-0.09}$  & $1.25^{+0.09}_{-0.07}$  & $1.41^{+1.87}_{-1.60}$  & $2.06^{+1.70}_{-1.93}$  & $0.08^{+0.03}_{-0.03}$  & $1.19^{+1.05}_{-1.11}$  & $3.57^{+1.82}_{-1.89}$ \\ 
$L_{10}$ & $-0.12^{+0.10}_{-0.08}$  & $-9.91^{+0.19}_{-0.25}$  & $1.54^{+0.17}_{-0.15}$  & $0.42^{+0.06}_{-0.06}$  & $0.24^{+0.03}_{-0.03}$  & $0.81^{+0.10}_{-0.09}$  & $1.25^{+0.09}_{-0.07}$  & $1.61^{+1.64}_{-1.50}$  & $1.99^{+1.47}_{-1.44}$  & $0.08^{+0.03}_{-0.03}$  & $1.31^{+0.89}_{-0.94}$  & $3.65^{+1.51}_{-1.83}$ \\ 
$L_{11}$ & $-0.09^{+0.10}_{-0.09}$  & $-9.91^{+0.18}_{-0.23}$  & $1.56^{+0.17}_{-0.15}$  & $0.43^{+0.07}_{-0.06}$  & $0.27^{+0.03}_{-0.03}$  & $0.85^{+0.11}_{-0.09}$  & $1.26^{+0.09}_{-0.07}$  & $1.53^{+1.46}_{-1.23}$  & $2.03^{+1.21}_{-1.32}$  & $0.09^{+0.03}_{-0.03}$  & $1.41^{+0.77}_{-0.78}$  & $3.55^{+1.37}_{-1.46}$ \\ 
$L_{12}$ & $-0.04^{+0.08}_{-0.08}$  & $-9.94^{+0.18}_{-0.20}$  & $1.59^{+0.16}_{-0.17}$  & $0.45^{+0.06}_{-0.06}$  & $0.29^{+0.03}_{-0.04}$  & $0.91^{+0.10}_{-0.10}$  & $1.26^{+0.09}_{-0.07}$  & $1.51^{+1.33}_{-1.15}$  & $2.03^{+1.13}_{-1.03}$  & $0.09^{+0.03}_{-0.02}$  & $1.61^{+0.66}_{-0.66}$  & $3.31^{+1.26}_{-1.48}$ \\ 
$L_{13}$ & $0.06^{+0.08}_{-0.08}$  & $-10.02^{+0.21}_{-0.20}$  & $1.72^{+0.15}_{-0.17}$  & $0.51^{+0.05}_{-0.06}$  & $0.33^{+0.04}_{-0.05}$  & $1.02^{+0.09}_{-0.11}$  & $1.29^{+0.12}_{-0.08}$  & $1.77^{+1.38}_{-1.13}$  & $2.13^{+0.97}_{-0.95}$  & $0.10^{+0.04}_{-0.03}$  & $1.91^{+0.65}_{-0.66}$  & $2.92^{+1.26}_{-1.54}$ \\ 
$L_{14}$ & $0.20^{+0.11}_{-0.09}$  & $-10.19^{+0.26}_{-0.22}$  & $1.94^{+0.17}_{-0.17}$  & $0.60^{+0.05}_{-0.05}$  & $0.37^{+0.05}_{-0.05}$  & $1.18^{+0.12}_{-0.12}$  & $1.35^{+0.13}_{-0.10}$  & $2.35^{+1.38}_{-1.24}$  & $2.37^{+0.93}_{-0.96}$  & $0.12^{+0.04}_{-0.04}$  & $2.29^{+0.79}_{-0.70}$  & $2.35^{+1.48}_{-1.54}$ \\ 
$L_{15}$ & $0.29^{+0.15}_{-0.14}$  & $-10.50^{+0.27}_{-0.29}$  & $2.15^{+0.21}_{-0.21}$  & $0.66^{+0.08}_{-0.08}$  & $0.38^{+0.07}_{-0.06}$  & $1.28^{+0.19}_{-0.16}$  & $1.44^{+0.17}_{-0.13}$  & $3.17^{+1.32}_{-1.57}$  & $2.65^{+0.99}_{-1.22}$  & $0.13^{+0.05}_{-0.04}$  & $2.38^{+0.94}_{-0.90}$  & $1.77^{+1.95}_{-1.74}$ \\ 
$L_{16}$ & $0.23^{+0.18}_{-0.15}$  & $-10.75^{+0.32}_{-0.39}$  & $2.15^{+0.26}_{-0.25}$  & $0.65^{+0.09}_{-0.08}$  & $0.35^{+0.08}_{-0.05}$  & $1.20^{+0.23}_{-0.18}$  & $1.49^{+0.19}_{-0.15}$  & $3.67^{+1.56}_{-1.67}$  & $2.87^{+1.02}_{-1.38}$  & $0.13^{+0.05}_{-0.04}$  & $1.99^{+1.04}_{-0.93}$  & $1.50^{+2.10}_{-2.14}$ \\ 
$L_{17}$ & $0.20^{+0.13}_{-0.11}$  & $-11.11^{+0.22}_{-0.35}$  & $2.20^{+0.17}_{-0.13}$  & $0.66^{+0.05}_{-0.05}$  & $0.33^{+0.04}_{-0.03}$  & $1.18^{+0.12}_{-0.10}$  & $1.60^{+0.15}_{-0.17}$  & $4.47^{+1.43}_{-1.67}$  & $3.08^{+1.28}_{-1.36}$  & $0.15^{+0.04}_{-0.04}$  & $1.83^{+0.86}_{-0.80}$  & $0.43^{+1.98}_{-2.03}$ \\ 
$L_{18}$ & $0.27^{+0.09}_{-0.10}$  & $-11.58^{+0.27}_{-0.34}$  & $2.35^{+0.15}_{-0.13}$  & $0.69^{+0.05}_{-0.04}$  & $0.35^{+0.04}_{-0.03}$  & $1.24^{+0.09}_{-0.09}$  & $1.70^{+0.14}_{-0.14}$  & $4.97^{+1.01}_{-1.19}$  & $3.20^{+0.84}_{-0.99}$  & $0.17^{+0.04}_{-0.04}$  & $2.14^{+0.88}_{-0.73}$  & $-0.55^{+1.75}_{-1.40}$ \\ 
$L_{19}$ & $0.34^{+0.08}_{-0.07}$  & $-11.93^{+0.38}_{-0.31}$  & $2.48^{+0.13}_{-0.13}$  & $0.73^{+0.04}_{-0.04}$  & $0.38^{+0.03}_{-0.03}$  & $1.32^{+0.08}_{-0.07}$  & $1.83^{+0.11}_{-0.13}$  & $5.38^{+0.67}_{-0.82}$  & $3.34^{+0.61}_{-0.67}$  & $0.22^{+0.04}_{-0.04}$  & $2.94^{+0.77}_{-0.76}$  & $-1.41^{+1.35}_{-1.06}$ \\ 
$L_{20}$ & $0.38^{+0.08}_{-0.07}$  & $-12.19^{+0.41}_{-0.27}$  & $2.57^{+0.12}_{-0.12}$  & $0.76^{+0.04}_{-0.04}$  & $0.39^{+0.03}_{-0.02}$  & $1.37^{+0.08}_{-0.06}$  & $1.91^{+0.08}_{-0.11}$  & $5.51^{+0.45}_{-0.56}$  & $3.38^{+0.44}_{-0.48}$  & $0.25^{+0.03}_{-0.04}$  & $3.65^{+0.70}_{-0.71}$  & $-1.91^{+1.03}_{-0.72}$ \\ 
\tableline
Blob & $0.61^{+0.25}_{-0.18}$  & $-11.99^{+0.40}_{-0.37}$  & $2.62^{+0.21}_{-0.16}$  & $0.75^{+0.04}_{-0.04}$  & $0.38^{+0.03}_{-0.03}$  & $1.34^{+0.09}_{-0.06}$  & $1.85^{+0.11}_{-0.11}$  & $5.49^{+0.57}_{-1.07}$  & $3.43^{+0.46}_{-0.47}$  & $0.23^{+0.03}_{-0.04}$  & $3.18^{+0.86}_{-0.77}$  & $-1.73^{+1.32}_{-0.94}$ \\ 
Red & $0.40^{+0.03}_{-0.03}$  & $-12.36^{+0.14}_{-0.14}$  & $2.62^{+0.05}_{-0.06}$  & $0.77^{+0.02}_{-0.02}$  & $0.39^{+0.02}_{-0.02}$  & $1.38^{+0.04}_{-0.04}$  & $1.95^{+0.06}_{-0.06}$  & $5.59^{+0.38}_{-0.38}$  & $3.45^{+0.42}_{-0.42}$  & $0.26^{+0.02}_{-0.02}$  & $3.87^{+0.54}_{-0.55}$  & $-2.21^{+0.64}_{-0.59}$ \\ 
Spine & & & & & & & & & & & & \\

\tableline
\tableline
{\tiny \rm Group} & log OIII & {\tt eclass} & log SFR & $\mu_{50,r}$ & $M_{r}$ & log $R90_r$ & $\log M_{*}$ & $\log \mu_{*,r}$ & log H$\alpha$ & log NII & $R90_{r}/R50_{r}$ & {\tt fracDeV}$_r$\\ 
	  &	(\AA)	 & 			 & ($M_{\odot}$yr$^{-1}$)  &  ({\tiny mag asec$^{ -2}$}) & (Mag)  &  (kpc)  & ($M_{\odot}$)  & ($M_{\odot}$kpc$^{-2}$) & (\AA) & (\AA) &       &  \\ 
\tableline
\tableline
$L_{1}$ & $1.57^{+0.41}_{-0.41}$  & $0.44^{+0.17}_{-0.10}$  & $-0.51^{+0.46}_{-0.34}$  & $20.42^{+0.71}_{-0.59}$  & $-18.09^{+0.61}_{-1.15}$  & $0.45^{+0.23}_{-0.17}$  & $8.63^{+0.54}_{-0.36}$  & $8.35^{+0.35}_{-0.43}$  & $1.87^{+0.29}_{-0.27}$  & $0.89^{+0.28}_{-0.32}$  & $2.65^{+0.27}_{-0.23}$  & $0.55^{+0.37}_{-0.35}$ \\ 
$L_{2}$ & $1.35^{+0.38}_{-0.35}$  & $0.36^{+0.12}_{-0.09}$  & $-0.75^{+0.45}_{-0.30}$  & $20.85^{+0.86}_{-0.56}$  & $-17.87^{+0.62}_{-1.08}$  & $0.47^{+0.23}_{-0.17}$  & $8.64^{+0.45}_{-0.36}$  & $8.25^{+0.35}_{-0.46}$  & $1.68^{+0.26}_{-0.22}$  & $0.74^{+0.24}_{-0.24}$  & $2.53^{+0.24}_{-0.21}$  & $0.28^{+0.45}_{-0.22}$ \\ 
$L_{3}$ & $1.23^{+0.28}_{-0.32}$  & $0.32^{+0.09}_{-0.08}$  & $-0.82^{+0.41}_{-0.28}$  & $21.26^{+0.55}_{-0.48}$  & $-17.85^{+0.65}_{-0.95}$  & $0.51^{+0.21}_{-0.18}$  & $8.61^{+0.41}_{-0.32}$  & $8.09^{+0.29}_{-0.34}$  & $1.57^{+0.24}_{-0.21}$  & $0.61^{+0.25}_{-0.29}$  & $2.43^{+0.21}_{-0.21}$  & $0.14^{+0.25}_{-0.14}$ \\ 
$L_{4}$ & $1.12^{+0.29}_{-0.30}$  & $0.29^{+0.09}_{-0.07}$  & $-0.81^{+0.36}_{-0.24}$  & $21.76^{+0.51}_{-0.55}$  & $-17.85^{+0.49}_{-0.87}$  & $0.61^{+0.16}_{-0.17}$  & $8.61^{+0.37}_{-0.22}$  & $7.89^{+0.31}_{-0.31}$  & $1.49^{+0.22}_{-0.20}$  & $0.54^{+0.25}_{-0.24}$  & $2.33^{+0.24}_{-0.19}$  & $0.06^{+0.19}_{-0.06}$ \\ 
$L_{5}$ & $1.06^{+0.35}_{-0.31}$  & $0.26^{+0.10}_{-0.07}$  & $-0.89^{+0.34}_{-0.19}$  & $22.24^{+0.57}_{-0.60}$  & $-17.73^{+0.41}_{-0.79}$  & $0.66^{+0.13}_{-0.14}$  & $8.57^{+0.33}_{-0.21}$  & $7.70^{+0.29}_{-0.32}$  & $1.43^{+0.26}_{-0.22}$  & $0.48^{+0.25}_{-0.23}$  & $2.20^{+0.21}_{-0.18}$  & $0.02^{+0.14}_{-0.02}$ \\ 
$L_{6}$ & $0.82^{+0.40}_{-0.26}$  & $0.20^{+0.09}_{-0.08}$  & $-0.97^{+0.31}_{-0.20}$  & $22.52^{+0.54}_{-0.47}$  & $-17.71^{+0.41}_{-0.68}$  & $0.70^{+0.11}_{-0.15}$  & $8.61^{+0.26}_{-0.17}$  & $7.65^{+0.20}_{-0.22}$  & $1.27^{+0.26}_{-0.20}$  & $0.38^{+0.25}_{-0.22}$  & $2.12^{+0.17}_{-0.17}$  & $0.00^{+0.12}_{-0.00}$ \\ 
$L_{7}$ & $0.71^{+0.33}_{-0.32}$  & $0.16^{+0.07}_{-0.08}$  & $-0.98^{+0.31}_{-0.29}$  & $22.53^{+0.47}_{-0.52}$  & $-17.88^{+0.48}_{-0.69}$  & $0.70^{+0.13}_{-0.13}$  & $8.76^{+0.25}_{-0.19}$  & $7.72^{+0.20}_{-0.21}$  & $1.20^{+0.21}_{-0.23}$  & $0.37^{+0.22}_{-0.22}$  & $2.07^{+0.21}_{-0.17}$  & $0.00^{+0.14}_{-0.00}$ \\ 
$L_{8}$ & $0.57^{+0.34}_{-0.37}$  & $0.14^{+0.08}_{-0.08}$  & $-0.97^{+0.37}_{-0.56}$  & $22.24^{+0.40}_{-0.46}$  & $-17.95^{+0.68}_{-0.84}$  & $0.67^{+0.16}_{-0.15}$  & $8.86^{+0.31}_{-0.27}$  & $7.89^{+0.16}_{-0.14}$  & $1.13^{+0.26}_{-0.25}$  & $0.42^{+0.22}_{-0.23}$  & $2.18^{+0.24}_{-0.22}$  & $0.01^{+0.17}_{-0.01}$ \\ 
$L_{9}$ & $0.60^{+0.32}_{-0.38}$  & $0.14^{+0.08}_{-0.08}$  & $-0.87^{+0.44}_{-0.34}$  & $21.91^{+0.40}_{-0.53}$  & $-18.22^{+0.66}_{-0.92}$  & $0.67^{+0.19}_{-0.16}$  & $9.00^{+0.34}_{-0.24}$  & $8.07^{+0.16}_{-0.14}$  & $1.21^{+0.22}_{-0.26}$  & $0.50^{+0.22}_{-0.26}$  & $2.23^{+0.25}_{-0.21}$  & $0.00^{+0.18}_{-0.00}$ \\ 
$L_{10}$ & $0.55^{+0.36}_{-0.37}$  & $0.14^{+0.09}_{-0.08}$  & $-0.68^{+0.48}_{-0.43}$  & $21.59^{+0.47}_{-0.54}$  & $-18.64^{+0.88}_{-1.02}$  & $0.70^{+0.20}_{-0.21}$  & $9.20^{+0.37}_{-0.32}$  & $8.22^{+0.19}_{-0.15}$  & $1.25^{+0.24}_{-0.24}$  & $0.62^{+0.23}_{-0.23}$  & $2.27^{+0.22}_{-0.23}$  & $0.03^{+0.21}_{-0.03}$ \\ 
$L_{11}$ & $0.50^{+0.39}_{-0.41}$  & $0.14^{+0.09}_{-0.08}$  & $-0.43^{+0.45}_{-0.50}$  & $21.26^{+0.51}_{-0.56}$  & $-19.19^{+1.05}_{-0.97}$  & $0.74^{+0.20}_{-0.24}$  & $9.44^{+0.37}_{-0.39}$  & $8.38^{+0.18}_{-0.15}$  & $1.31^{+0.23}_{-0.25}$  & $0.72^{+0.23}_{-0.25}$  & $2.28^{+0.23}_{-0.25}$  & $0.04^{+0.23}_{-0.04}$ \\ 
$L_{12}$ & $0.43^{+0.43}_{-0.43}$  & $0.13^{+0.11}_{-0.10}$  & $-0.24^{+0.44}_{-0.49}$  & $20.86^{+0.56}_{-0.63}$  & $-19.63^{+1.21}_{-0.99}$  & $0.75^{+0.21}_{-0.28}$  & $9.67^{+0.39}_{-0.49}$  & $8.59^{+0.21}_{-0.18}$  & $1.37^{+0.25}_{-0.28}$  & $0.83^{+0.24}_{-0.25}$  & $2.30^{+0.28}_{-0.24}$  & $0.07^{+0.38}_{-0.07}$ \\ 
$L_{13}$ & $0.29^{+0.50}_{-0.38}$  & $0.08^{+0.12}_{-0.10}$  & $-0.04^{+0.42}_{-0.50}$  & $20.58^{+0.56}_{-0.65}$  & $-20.09^{+1.28}_{-0.98}$  & $0.79^{+0.20}_{-0.29}$  & $9.95^{+0.41}_{-0.54}$  & $8.80^{+0.24}_{-0.21}$  & $1.36^{+0.27}_{-0.32}$  & $0.87^{+0.27}_{-0.29}$  & $2.35^{+0.29}_{-0.24}$  & $0.17^{+0.46}_{-0.17}$ \\ 
$L_{14}$ & $0.20^{+0.48}_{-0.32}$  & $0.00^{+0.10}_{-0.08}$  & $0.01^{+0.45}_{-0.48}$  & $20.46^{+0.62}_{-0.64}$  & $-20.30^{+1.28}_{-1.02}$  & $0.83^{+0.21}_{-0.27}$  & $10.18^{+0.44}_{-0.57}$  & $9.00^{+0.28}_{-0.27}$  & $1.27^{+0.27}_{-0.33}$  & $0.84^{+0.28}_{-0.29}$  & $2.47^{+0.31}_{-0.23}$  & $0.34^{+0.47}_{-0.33}$ \\ 
$L_{15}$ & $0.13^{+0.46}_{-0.33}$  & $-0.06^{+0.10}_{-0.08}$  & $-0.37^{+0.43}_{-0.50}$  & $20.62^{+0.75}_{-0.67}$  & $-19.90^{+1.39}_{-1.25}$  & $0.80^{+0.24}_{-0.26}$  & $10.13^{+0.55}_{-0.70}$  & $9.03^{+0.31}_{-0.36}$  & $1.05^{+0.28}_{-0.37}$  & $0.64^{+0.28}_{-0.29}$  & $2.55^{+0.32}_{-0.24}$  & $0.40^{+0.47}_{-0.36}$ \\ 
$L_{16}$ & $-0.03^{+0.44}_{-0.37}$  & $-0.06^{+0.09}_{-0.08}$  & $-1.08^{+0.49}_{-0.33}$  & $20.95^{+0.82}_{-0.82}$  & $-18.78^{+1.08}_{-1.81}$  & $0.67^{+0.27}_{-0.18}$  & $9.61^{+0.85}_{-0.60}$  & $8.85^{+0.41}_{-0.43}$  & $0.79^{+0.29}_{-0.68}$  & $0.38^{+0.25}_{-0.41}$  & $2.56^{+0.27}_{-0.25}$  & $0.40^{+0.44}_{-0.36}$ \\ 
$L_{17}$ & $-0.34^{+0.40}_{-0.50}$  & $-0.08^{+0.05}_{-0.05}$  & $-1.69^{+0.47}_{-0.35}$  & $21.13^{+0.71}_{-0.79}$  & $-18.16^{+0.70}_{-1.56}$  & $0.59^{+0.19}_{-0.13}$  & $9.30^{+0.72}_{-0.32}$  & $8.75^{+0.37}_{-0.38}$  & $-0.19^{+0.93}_{-0.42}$  & $-0.16^{+0.54}_{-0.81}$  & $2.54^{+0.22}_{-0.21}$  & $0.42^{+0.35}_{-0.28}$ \\ 
$L_{18}$ & $-0.41^{+0.42}_{-0.55}$  & $-0.11^{+0.04}_{-0.04}$  & $-1.77^{+0.45}_{-0.53}$  & $20.36^{+0.79}_{-0.57}$  & $-19.00^{+1.23}_{-1.30}$  & $0.60^{+0.21}_{-0.17}$  & $9.73^{+0.58}_{-0.56}$  & $9.12^{+0.24}_{-0.38}$  & $-0.35^{+0.67}_{-0.28}$  & $-0.49^{+0.69}_{-0.63}$  & $2.64^{+0.20}_{-0.17}$  & $0.71^{+0.24}_{-0.35}$ \\ 
$L_{19}$ & $-0.31^{+0.30}_{-0.55}$  & $-0.14^{+0.03}_{-0.03}$  & $-1.54^{+0.42}_{-0.38}$  & $19.68^{+0.49}_{-0.47}$  & $-20.39^{+1.03}_{-0.78}$  & $0.73^{+0.22}_{-0.21}$  & $10.35^{+0.36}_{-0.44}$  & $9.46^{+0.18}_{-0.21}$  & $-0.21^{+0.44}_{-0.37}$  & $-0.29^{+0.48}_{-0.74}$  & $2.84^{+0.17}_{-0.16}$  & $0.96^{+0.04}_{-0.17}$ \\ 
$L_{20}$ & $-0.27^{+0.26}_{-0.42}$  & $-0.15^{+0.03}_{-0.03}$  & $-1.35^{+0.35}_{-0.31}$  & $19.41^{+0.50}_{-0.47}$  & $-21.32^{+0.73}_{-0.72}$  & $0.93^{+0.20}_{-0.20}$  & $10.77^{+0.31}_{-0.32}$  & $9.60^{+0.19}_{-0.21}$  & $-0.11^{+0.36}_{-0.44}$  & $-0.17^{+0.41}_{-0.62}$  & $3.18^{+0.18}_{-0.17}$  & $1.00^{+0.00}_{-0.04}$ \\ 
\tableline
Blob & $-0.56^{+0.38}_{-0.38}$  & $-0.14^{+0.03}_{-0.02}$  & $-1.68^{+0.35}_{-0.38}$  & $18.83^{+0.29}_{-0.31}$  & $-19.37^{+1.10}_{-1.08}$  & $0.21^{+0.24}_{-0.18}$  & $10.31^{+0.47}_{-0.68}$  & $10.06^{+0.38}_{-0.30}$  & $-0.40^{+0.47}_{-0.22}$  & $-0.48^{+0.53}_{-0.41}$  & $1.95^{+0.15}_{-0.11}$  & $0.00^{+0.48}_{-0.00}$ \\ 
Red & $-0.27^{+0.24}_{-0.36}$  & $-0.16^{+0.02}_{-0.02}$  & $-1.45^{+0.11}_{-0.11}$  & $19.30^{+0.19}_{-0.20}$  & $-21.52^{+0.48}_{-0.45}$  & $0.96^{+0.10}_{-0.10}$  & $10.85^{+0.20}_{-0.20}$  & $9.65^{+0.10}_{-0.09}$  & $-0.16^{+0.30}_{-0.38}$  & $-0.24^{+0.38}_{-0.54}$  & $3.23^{+0.10}_{-0.09}$  & $1.00^{+0.00}_{-0.00}$ \\ 
Spine & & & & & & & & & & & & \\
\tableline
\tableline
\label{Table:20GroupAveProp}
\end{tabular}
\end{center}
{The blob and red spine groups are detailed in Sec. \ref{Sec:InterestingGroups}.}
\end{table*}

\subsection{Spectral and Physical Properties}\label{Sec:SpecAndPhysProp}

Figure \ref{Fig:LambdaVsProp} shows the evolution of the galaxy properties as a function of arc length in the principal curve, whereas Table \ref{Table:20GroupAveProp} contains the average values of the properties at each $\lbrace L_{i}\rbrace_{i=1}^{i=20}$.

Looking at the 7 properties on which the WPCA was built, the most important feature is that they present the same shape as the $\mathbf{PC}_{i}$ in which they have the greatest leverage. 
On the other hand, the properties not present in the WPCA show similar shapes or behaviors, depending to their individual correlations with the initial 7 properties. Generally, their behavior (as a function of arc-length) is defined by the 4 branches and 3 turning points, resembling in most cases a distorted $W$ of the P-curve.

Thus, we can group all the properties with respect to which $\mathbf{PC}_{i}$ they resemble the most. For example, Fig. \ref{Fig:PCAEigenvectors} shows that $\log M_{*}/{\rm L}_{r}$, $\log SFR/M_{*}$, $u-r$ and $g-r$ have the greatest leverage in $\mathbf{PC}_{1}$ from all first 4 PCs. This makes these properties resembling like the shape of $\mathbf{PC}_{1}$ in Fig. \ref{Fig:PCAvsLambda}, where it's mostly a linear relation with respect to arc length, with a scatter modulated by the turning points. Other properties that correlate with $\mathbf{PC}_{1}$ are $r-i$, $g-z$, $D_{n}(4000)$, some Lick indices and [OIII]. Note that [OIII] has a strong linear dependence on $\mathbf{PC}_{1}$ and behaves differently to the other emission lines (such as Balmer series) due to the higher ionization degree. The arc length also correlates linearly with {\tt eclass}, which is the classification parameter derived in the PCA of galaxy spectra in \cite{yip2004b}, defined as a function of the expansion coefficients in the base of the first 2 eigenspectra. In general, these properties can be expressed as a linear combination of each other, as seen in astrophysical use, e.g. $M_{*}$/L = $a\times$ Color + $b$ \citep[e.g.][]{baldry2004}.

In the same way, $\log SFR$ and $\mu_{50,r}$ resemble strongly the "W" shape of $\mathbf{PC}_{2}$. Some emission lines belonging this group are NII and H$_{\alpha}$, the latter being a well known proxy for $SFR$ \citep{kennicutt1998}. Also, $M_{r}$ is a good proxy for $\log M_{*}$ \citep{bell2003}, both of which related to $R90_{r}$ and $\mu_{*,r}$, and seemingly correlate with $\mu_{50,r}$. Interestingly, the shape of the average Lick Na D index is similar to the "W" shape of $\mathbf{PC}_{2}$ (and $SFR$), but the larger 1$\sigma$ dispersion in the average makes it not very significant. Lick Na D is expected to represent a strong absorption feature in old stellar populations, as shown in the ellipticals at high arc lengths. We can however see that it presents also a relatively high average at the second turning point. This is related to the fact that Na absorption is not only present in stars, but also in the inter stellar medium as a consequence of outflows or winds present in high star-formation spiral galaxies \citep{chen2010}.

Note that the boomerang shape of $R90_{r}/R50_{r}$ is almost identical to $\mathbf{PC}_{3}$. Also correlating with the concentration index is {\tt fracDeV}$_{r}$ \citep{stoughton2002}, which determines the mixing in the modeling of the light profiles galaxies, between an exponential disk and a de Vaucouleurs $r^{1/4}$ law for the elliptical bulge.

\subsubsection{The BPT Diagram}\label{Sec:TheBPTdiagram}

\begin{figure}
\epsscale{1.0}
\plotone{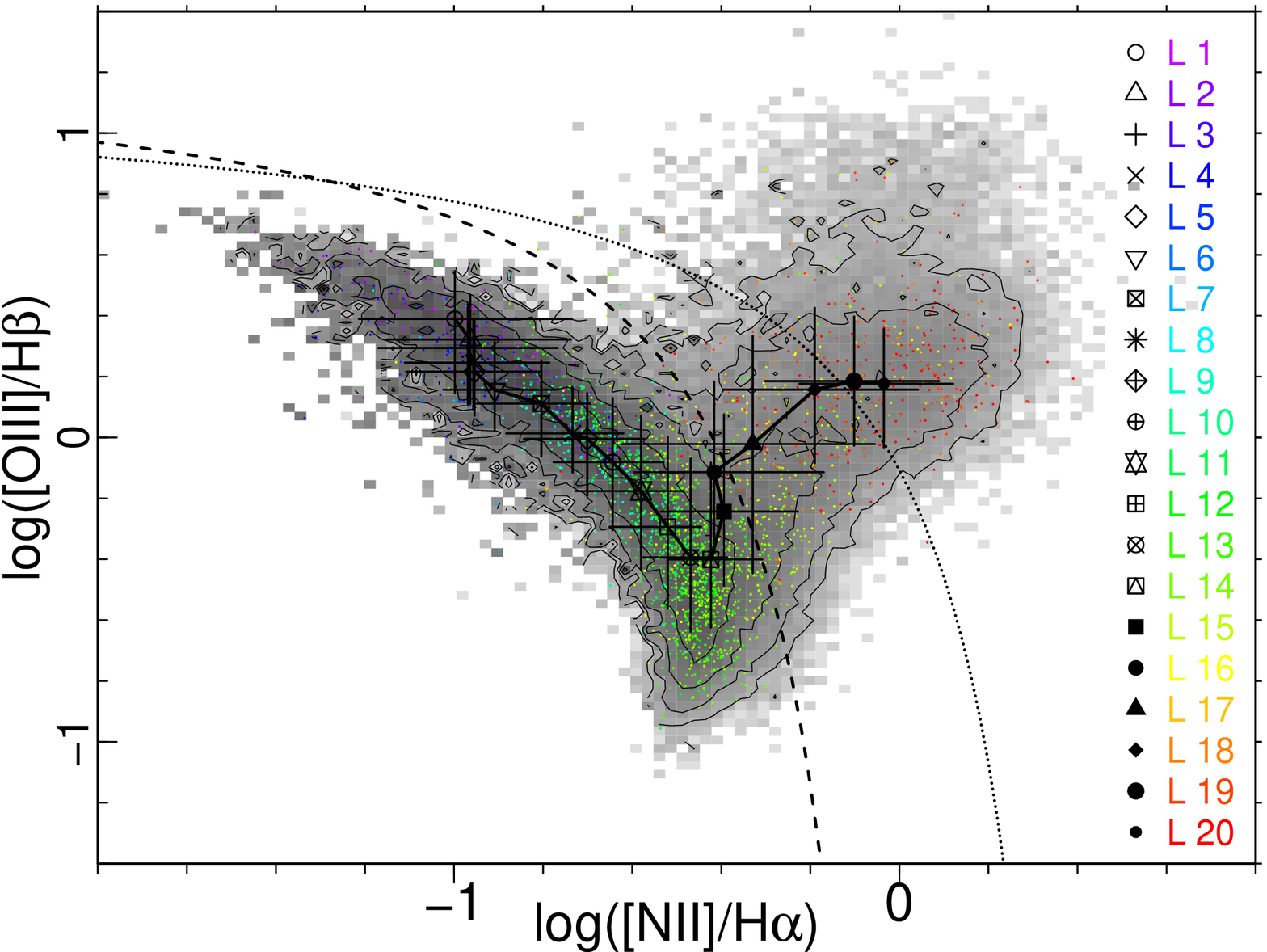}
\caption{BPT diagram of the MGS galaxy sample. Symbols connected with straight lines track the positions of the average $\log$[NII]/H$_{\alpha}$ and $\log$[OIII]/H$_{\beta}$ of each $\lbrace L_{i}\rbrace_{i=1}^{i=20}$ ($1\sigma$ dispersion bars also included). Colored dots are a random 2\% samples of each group. Dashed and dotted lines show the separators from \cite{kauffmann2003b} and \cite{kewley2001}, respectively, between pure starforming galaxies (left region), composite (central) and AGN (right).} 
\label{Fig:BPTdiagram}
\end{figure}

Figure \ref{Fig:BPTdiagram} shows the emission-line ratios BPT diagram \citep{baldwin1981} of the MGS galaxies, where AGN identification can be done easily. We considered galaxies presenting H$\alpha$, H$\beta$, [NII] and [OIII] emission lines with well measured equivalent widths, of fractional errors smaller than 0.33 and velocity dispersion smaller than 500km s$^{-1}$. These cuts make the $L_{1}-L_{15}$ groups be reduced to $\sim$90\% of their size, going down to $~35$\% for the remaining groups at higher arc length. Since none of these emission lines where included in the building of the WPCA, we overplotted the average location of the $\lbrace L_{i}\rbrace_{i=1}^{i=20}$ groups. 

The average locations of the groups can be seen to be connected by a two-branched track in line-ratio space. The left branch covers the region of star forming galaxies, whereas the right branch crosses the separator of \cite{kauffmann2003b} into the region filled by AGNs. Interestingly, the joining point between the 2 branches happens at $L_{14}$, which contains the 2nd turning point in the principal curve. This is a striking feature, as it shows that the P-curve is powerful enough to describe galaxy properties beyond the ones included in its construction.

\begin{figure*}
\epsscale{1.0}
\plotone{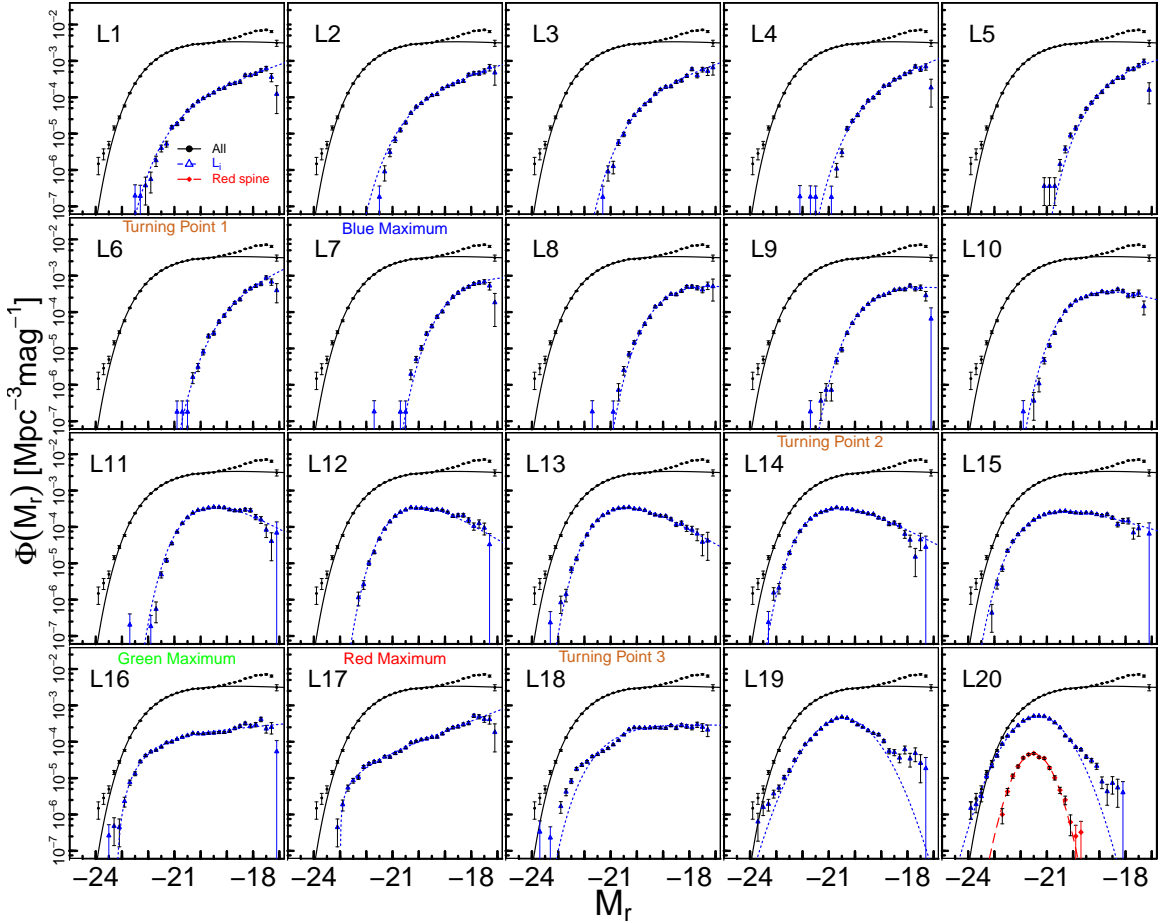}
\caption{Luminosity functions of the $\lbrace L_{i}\rbrace_{i=1}^{i=20}$ groups ordered by increasing arc-length (blue triangles with a dashed line fit from Table \ref{Table:FittingParametersLumFun}). The aggregated luminosity function of all the $L_{i}$ samples is shown as black circles with the Schechter fit as a black continuous line. The red diamonds denote the luminosity function belonging to the group of red galaxies located very close to the principal curve within the $L_{20}$ group (see Sec. \ref{Sec:RedSpineGaussianLumFun}).}
\label{Fig:LumFun20sep}
\end{figure*}

\begin{table}[h]
\begin{center}
\caption{Luminosity functions fitting parameters.\rlap{$^a$}}
\scriptsize \addtolength{\tabcolsep}{-5pt}
\begin{tabular}{ccccc}
\tableline
Group & $\phi^{*} \times 10^3$ \rlap{$^b$}  & $M^{*}$ & $\alpha$ & $\xi$\\ 
\tableline
All & $4.90\pm 0.14 $  & $-21.30\pm 0.03 $  & $-0.91\pm 0.02 $  &  0 \\ 
\tableline
$L_{1}$ & $0.13\pm 0.02 $  & $-20.45\pm 0.12 $  & $-1.60\pm 0.05 $  &  0 \\ 
$L_{2}$ & $0.19\pm 0.03 $  & $-19.90\pm 0.13 $  & $-1.54\pm 0.07 $  &  0 \\ 
$L_{3}$ & $0.27\pm 0.08 $  & $-19.48\pm 0.21 $  & $-1.56\pm 0.13 $  &  0 \\ 
$L_{4}$ & $0.36\pm 0.08 $  & $-19.21\pm 0.15 $  & $-1.62\pm 0.10 $  &  0 \\ 
$L_{5}$ & $0.85\pm 0.21 $  & $-18.45\pm 0.18 $  & $-1.34\pm 0.18 $  &  0 \\ 
$L_{6}$ & $0.68\pm 0.19 $  & $-18.54\pm 0.17 $  & $-1.69\pm 0.15 $  &  0 \\ 
$L_{7}$ & $1.50\pm 0.10 $  & $-18.07\pm 0.07 $  & $-0.87\pm 0.10 $  &  0 \\ 
$L_{8}$ & $1.13\pm 0.09 $  & $-18.34\pm 0.11 $  & $-0.65\pm 0.16 $  &  0 \\ 
$L_{9}$ & $0.93\pm 0.05 $  & $-18.86\pm 0.06 $  & $-0.76\pm 0.07 $  &  0 \\ 
$L_{10}$ & $0.96\pm 0.04 $  & $-19.20\pm 0.06 $  & $-0.42\pm 0.08 $  &  0 \\ 
$L_{11}$ & $1.04\pm 0.01 $  & $-19.41\pm 0.04 $  & $0.03\pm 0.06 $  &  0 \\ 
$L_{12}$ & $1.00\pm 0.01 $  & $-19.85\pm 0.04 $  & $0.12\pm 0.06 $  &  0 \\ 
$L_{13}$ & $0.99\pm 0.01 $  & $-20.34\pm 0.02 $  & $0.06\pm 0.03 $  &  0 \\ 
$L_{14}$ & $0.95\pm 0.01 $  & $-20.68\pm 0.02 $  & $-0.07\pm 0.02 $  &  0 \\ 
$L_{15}$ & $0.69\pm 0.01 $  & $-20.92\pm 0.03 $  & $-0.46\pm 0.03 $  &  0 \\ 
\tableline
$L_{16}$ & $0.16\pm 0.02 $  & $-22.01\pm 0.15 $  & $-1.16\pm 0.04 $  & $-0.32\pm 0.02 $ \\ 
$L_{17}$ & $0.016\pm 0.004 $  & $-23.15\pm 0.28 $  & $-1.69\pm 0.03 $  & $-1.167\pm 0.004 $ \\ 
\tableline
$L_{18}$ & $0.41\pm 0.09 $  & $-20.69\pm 0.24 $  & $-0.93\pm 0.14 $  &  0 \\ 
\tableline
\tableline
Group & $\phi^{*} \times 10^4$ \rlap{$^b$} & $\mu_{M}$ & $\sigma_{M}$ & - \\ 
\tableline
$L_{19}$ & $8.52\pm 0.19 $  & $-20.46\pm 0.02 $  & $0.78\pm 0.02 $  &  - \\ 
$L_{20}$ & $9.00\pm 0.11 $  & $-21.34\pm 0.01 $  & $0.70\pm 0.01 $  &  - \\ 
Red Spine & $0.55\pm 0.02 $  & $-21.51\pm 0.02 $  & $0.46\pm 0.01 $  &  - \\ 
\tableline
\tableline
\label{Table:FittingParametersLumFun}
\end{tabular}
\end{center}
{\small
$^a$~ Fittings to Eqns. \ref{Eq:LumFunDoublePowLaw} and \ref{Eq:LumFunLogNormal}. The parameters in Magnitude-space can be expressed as luminosities using $M_{r}$=$-2.5\log_{\rm 10} [{\rm L}/{\rm L}_{\odot}]$+$M_{\odot,r}$, where $M_{\odot,r}=4.62$ \citep{blanton2001}. \\ $^b$~ $\phi^{*}$ in units of Mpc $^{-3}$ Mag $^{-1}$.  }
\end{table}

\subsection{Luminosity Functions}\label{Sec:LumfunVsArcLength}

Figure \ref{Fig:LumFun20sep} shows the luminosity function (LFs) corresponding to the $\lbrace L_{i} \rbrace _{i=1}^{i=20}$ equal number density groups. They can be directly compared to the evolution of $M_{r}$ as a function of arc-length in Fig. \ref{Fig:LambdaVsProp}. The LFs were computed with the $V_{\rm max}$ method of \cite{schmidt1968} explained in Sec \ref{Sec:Vmax}, where the estimated LF value at each magnitude bin is the sum of the weights of all galaxies in that bin, with $w_{i}=V_{{\rm max},i}^{-1}$. Table \ref{Table:FittingParametersLumFun} contains the fitting parameters, choosing the fitting functions to be: 

\begin{itemize}

\item 
Double power-law:
\begin{equation}
\Phi({\rm L})d{\rm L} = \phi^{*} \left( \frac{{\rm L}}{{\rm L}_{*}}\right)^{\alpha} \left( 1+ \xi  \left(\frac{{\rm L}}{{\rm L}_{*}}\right)   \right)^{-\frac{1}{\xi}}  \frac{d{\rm L}}{{\rm L}_{*}},
\label{Eq:LumFunDoublePowLaw} 
\end{equation}
granted that $1+ \xi  ({\rm L}/{\rm L}_{*})\geq 0$. This double power-law fit \citep{alcaniz2004} collapses when $\xi=0$ into the Schechter fit \citep{schechter1976}  $\Phi({\rm L})d{\rm L} = \phi^{*} \left( {\rm L}/{\rm L}_{*} \right)^\alpha \exp ( -{\rm L}/{\rm L}_{*}) d{\rm L}/{\rm L}_{*}$.  The $\xi$ parameter can be related to the tail index in extreme value statistics \citep[e.g][]{gumbel1958,galambos1978}, defining for the brightest luminosities an infinite reaching power law tail ($\xi > 0$), exponential tail ($\xi = 0$) or a cutoff at a finite maximum luminosity ${\rm L}_{max}={\rm L}_{*}/|\xi|$ ($\xi < 0$).

\item 

Log-normal:
\begin{equation}
\Phi({\rm L})d{\rm L} =  \frac{1}{ {\rm L} \sigma_{\rm L} \sqrt{2\pi} } \exp( -\frac{ (\ln {\rm L} - \mu_{\rm L}  )^2}{ 2\sigma_{\rm L}^2 }      )d{\rm L} 
\label{Eq:LumFunLogNormal} 
\end{equation}
Note that a Log-normal distribution in luminosity-space  is equivalent as a normal-Gaussian distribution in magnitude-space. 
\end{itemize}

The luminosity function of the whole sample is well fitted by the Schechter fit, except for the bumps at the high luminosity tail and at the low luminosity end starting at $M_{r} \sim -18.8$, also observed by \cite{blanton2003} and \cite{blanton2005}. We fitted it right before the bump. 

The changes in the behavior of the luminosity function along the P-curve are determined also by the 3 turning points. In summary, $M^{*}$ becomes fainter in the 1st branch, and then brighter afterward. On the other hand, the slope of the faint tail behaves similarly to $\mathbf{PC}_{2}$. In fact, it is steep in the 1st branch, and becomes shallower in the 2nd branch, then again steep in the 3rd branch. The 4th branch contains an extremely shallow slope, where the luminosity functions resemble mostly a log-normal distribution.
We can see that we recover the luminosity function shapes shown in \cite{binggeli1988}, which are based on morphological types, and vary between schechter fits (as in $L_{1}$ to $L_{15}$) and bell shaped luminosity functions fitted by Log-normal fits (such as the $L_{19}$ and $L_{20}$).

As we progress along the 1st branch of the principal curve ($L_{1}$ to $L_{6}$), the blue compact dwarfs become less luminous on average, having $M^{*}$ dimmer in about $\Delta M^{*} \simeq 2$, with a mostly constant steep faint-end ($\alpha \sim -1.55$). Note that these galaxies, and mostly the low surface brightness spirals at the 1st turning point, create the bump seen in the overall luminosity function. In fact, $L_{6}$ contains the faintest galaxies in our sample (Fig. \ref{Fig:LambdaVsProp}). At this point, $\alpha=-1.69$ gives the steepest power law slope at the faint luminosity tail. Note that this slope is expected to reach $\alpha \simeq -1.5$, as noted in \cite{blanton2005}.

In the second branch ($L_{7}$ to $L_{13}$), the star forming spirals present a faint-end that flattens dramatically and starts to drop continuously, with an increase of $\Delta \alpha \sim 1.6 $ from $L_{7}$. At the same time, they start becoming much more luminous, with $M^{*}$ brightening in $\Delta M^{*}\sim -2.2$ (from $L_{6}$).

In the 3rd branch ($L_{14}$ to $L_{17}$), for the red spirals and lenticulars $M^{*}$ continues becoming brighter ($\Delta M^{*} \sim -2.5$), but at the same time the faint-end slope starts turning steep again ($\Delta \alpha \sim -1.6$), back to the values of  $\alpha \sim -1.6$ found at the end of the 1st branch. Note that $L_{16}$ (green maximum) and $L_{17}$ (red maximum) show long power-law faint ends with a sharp cutoff at the bright end. They are better fitted by a double power law fit, and since $\xi<0$ they present bright-end finite cuts at $M_{r}\sim -23.0$ and $-23.3$ respectively.

The 3rd turning point ($L_{17}$) is a unique case. The luminosity function presents 3 power-law-like sections, the faintest one being flat. We attempted to fit it with a Schechter profile. 

In the last 2 groups ($L_{19}$ and $L_{20}$), the faint end tail has dropped enormously. We attempted a Log-normal fits for the luminosities, since the luminosity functions look more bell-shaped, specially the ones belonging to the $\lbrace\lambda_{i}\rbrace _{i=1}^{i=5}$ groups that track the spine.


\begin{figure*}
\epsscale{1.15}
\plottwo{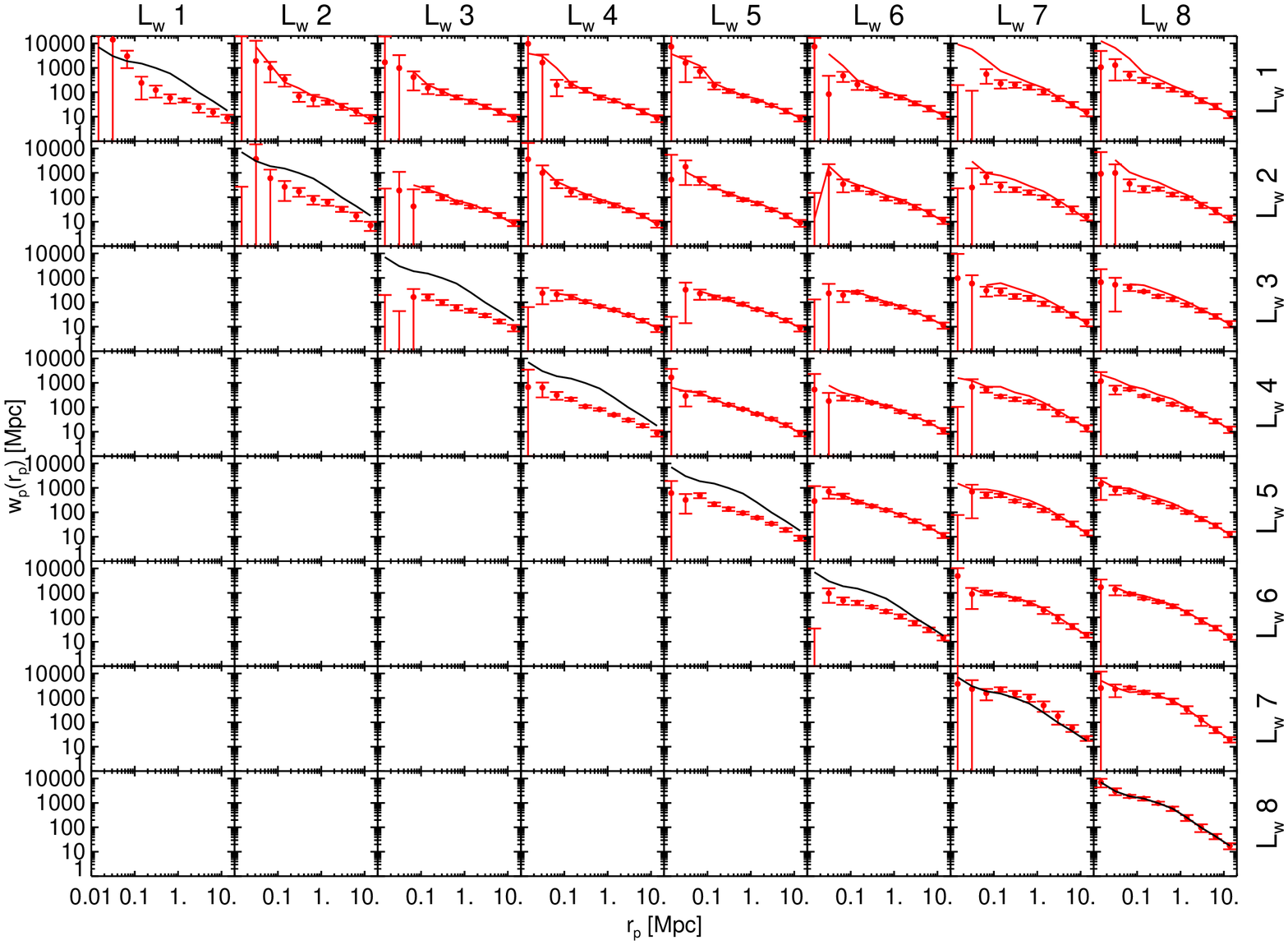}{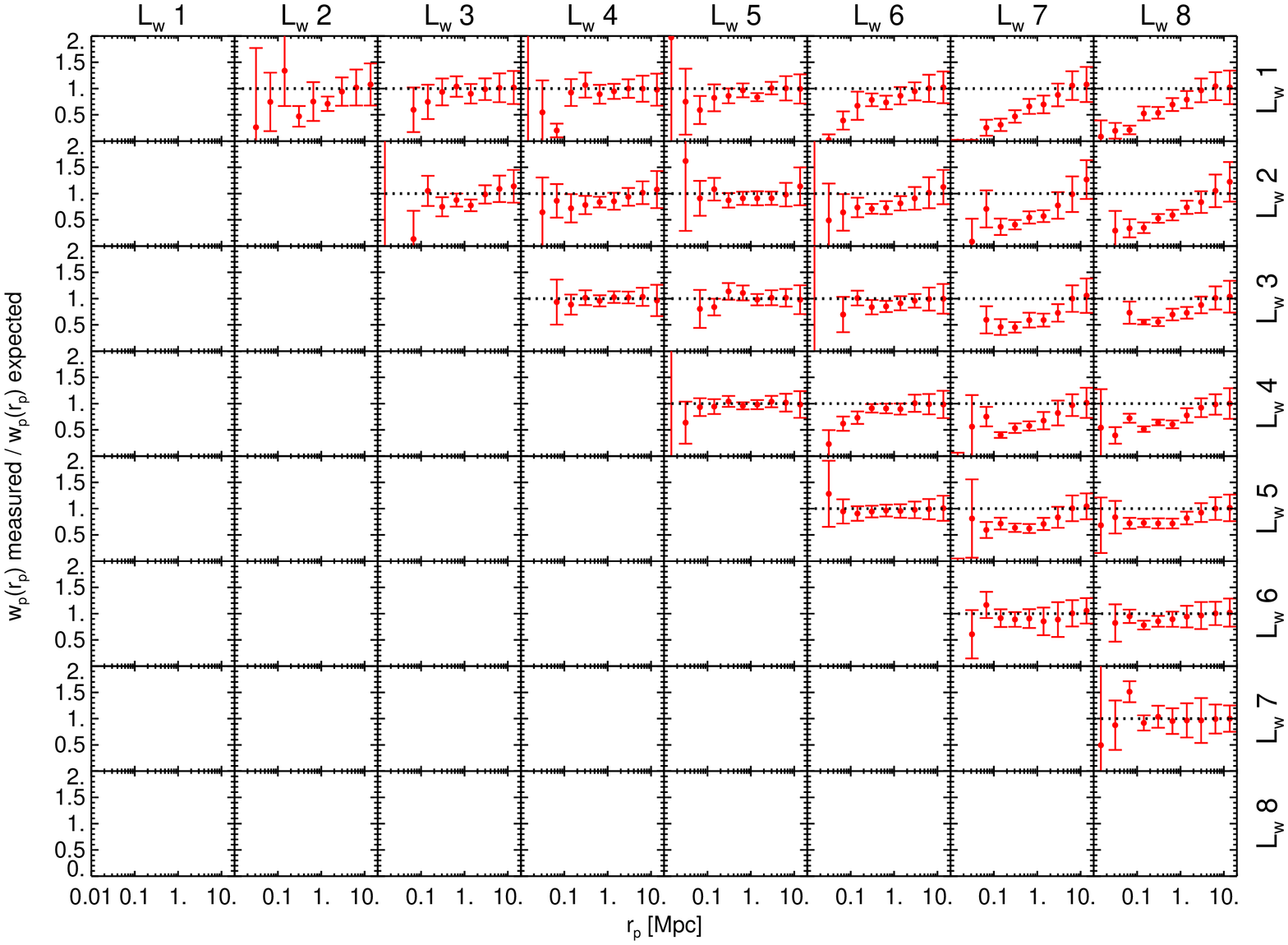}
\caption{\emph{Left:} The diagonal panels show the auto correlation function of the $\lbrace L_{w}i \rbrace$ groups. The solid black line is the auto correlation function of $L_{w}8$ shown for reference. Off diagonal panels show the cross correlation functions between the groups, where the red solid line represents the expected cross correlation function when the galaxies from the 2 samples are well mixed in the dark matter halos. \\ 
\emph{Right:} Same layout as in figure on the left, but showing the ratio between the measured cross correlation function to the expected one when the galaxies from the 2 samples are well mixed in the dark matter halos.}
\label{Fig:CorrFuncions} 
\end{figure*}

\subsection{Galaxy clustering}\label{Sec:Clustering}

In this section, we investigate the second moment of the galaxy
distribution as a function of the arc length, the spatial
distribution, quantified by the clustering. We explore here not only
the dependence of the galaxy clustering as a function of $L$, but also
the relative distribution of galaxies as a function of $L$, which can
be quantified by the cross-correlation function.

Following common practice, we compute first the redshift space
correlation function, as a function of the distances parallel ($\pi$)
and perpendicular ($r_p$) to the line of sight.  We use a generalized
version \citep{Szapudi_1998} of the \citet{Landy_1993} estimator

\begin{equation}\label{eq_xi}
  \xi(r_p, \pi) = \frac{D_aD_b - D_aR_b - D_bR_a + R_aR_b}{R_aR_b}
\end{equation}

where the subscripts $a$ and $b$ refer to the two samples we are
considering when measuring the cross-correlation function. We use the
same methods than \citet{Heinis_2009} to compute the correlation
functions. In brief, for each sample we generate random catalogs
following the SDSS footprint defined by its sectors. We use 50 times
more random objects than galaxies. We reproduce the selection function
by randomly drawing redshifts from the current sample. We correct from
fiber collision using the method described in
\citet{Heinis_2009}. Note that the fiber collision correction applies
only to the $D_aD_b$ term in Eq. \eqref{eq_xi}.

As $\xi(r_p, \pi)$ is sensitive to redshift distortions, we consider
the projected spatial correlation function afterward, which is free
from such effects:

\begin{equation}
  w_p(r_p) = 2\int_0^{\pi_{max}}\xi(r_p, \pi)d\pi
\end{equation}

where we use $\pi_{max} = 25$Mpc, for convergence purposes.

We compute error bars on $w_p(r_p)$ from jackknife resampling. We
build jackknife samples using the SDSS stripes, which are defined to
be 2.5 degrees wide great circles on the sky, following the survey
latitude. In practice we consider 23 jackknife samples built from the
stripes.

We use a volume limited sample extracted from our main sample (see
Sec. \ref{Sec:GalaxySample}). In order to maximize the signal-ratio of the
clustering measurements, we do not use 20 samples in $L$, but 8 of
them built the following way: we collided $L_1$ to $L_6$ in two groups
of similar number of galaxies ($L_{w}1$ and $L_{w}2$ ) , $L_7$ to $L_{10}$ in one group ($L_{w}3$), and for
the remaining $L$ samples, we grouped them two by two ($L_{w}4=\lbrace L_{11},L_{12} \rbrace$, $L_{w}5=\lbrace L_{13},L_{14} \rbrace$, $L_{w}6=\lbrace L_{15},L_{16} \rbrace$, $L_{w}7=\lbrace L_{17},L_{18} \rbrace$ and $L_{w}8=\lbrace L_{19},L_{20} \rbrace$).

Figure \ref{Fig:CorrFuncions} shows the results for the auto correlation functions
of these sample in the diagonal plots. As a reference, we show as
solid line in all diagonal plots the auto correlation function of the
sample with highest arc length ($L_{w}8$). The results from the auto correlation
function show that the amplitude of the correlation function at large
scales ($r_p\sim 10$ Mpc) increases with the arc length. This implies
that the host halo mass also increases with $l$. This result is
expected as $l$ does correlate with $u-r$ and $g-r$ colors for
instance. It is indeed well known that the amplitude of the
correlation function increases for redder objects in the local
Universe \citep[e.g.][]{Zehavi_2005, Zehavi_2011}. There are also some
interesting features in the small scales ($r_p < 0.1$ Mpc) clustering.
Indeed, most samples show clustering power at all scales, except our
groups 2 and 3 in particular, where there is an indication of a lack
of pairs at $r_p<0.1$ Mpc, suggesting a population mainly composed of
central galaxies.

In the off-diagonal plots, we show the cross-correlation functions
between these samples. It is beyond the scope of this paper to fully
interpret all these measurements with Halo Occupation Distribution
models \citep[e.g.][]{Cooray_2002}. We will use simple arguments to
highlight the information contained in the cross-correlation function.

We represent as a solid line in all off-diagonal plots the expected
cross-correlation function given by

\begin{equation}\label{eq_wx}
  w_{X_{ab}}(r_p) = \sqrt{w_a(r_p) w_b(r_p)}
\end{equation}
where $w_a(r_p)$ and $w_b(r_p)$ are the autocorrelation functions of
samples $a$ and $b$. Eq. \eqref{eq_wx} gives the cross-correlation
function which is expected in the case where galaxies from the two
samples are well mixed in the dark matter halos hosting them
\citep[see e.g.][]{Zehavi_2005}. This is of interest at small scales,
where the cross-correlation function contains information about close
pairs of galaxies that lie within the same dark matter halos.  Our
results show an interesting trend in this context. Indeed the cross
correlation function of galaxies with close arc lengths is similar to
the expected cross correlation function from Eq. \eqref{eq_wx}. On the
other hand, the cross correlation of galaxies more distant in terms of
arc length diverges from the expected correlation function. In
particular, the measures are \textit{overestimated} by Eq \ref{eq_wx}
at scales $r_p <0.1$ Mpc. This means that there are less close galaxy
pairs in the measures than what is expected in the case of a perfect
mix between galaxies of different arc length. Note that there is still
clustering signal at these scales, which means that there are some
galaxies belonging to distant $L$ groups in the same dark matter
halos. However, our results show that there are fewer pairs than what
is expected if galaxies are properly mixed. This suggests that the
probability that two galaxies reside in the same dark matter halo
decreases aa a function of their distance in arc length.


\subsection{Interesting Galaxy Groups found from WPCA and Principal curve classification}\label{Sec:InterestingGroups}

The analysis of galaxy properties with WPCA and P-curve methods allowed us to find and pinpoint some relevant groups that stand aside from the main trends of the whole galaxy sample. In particular, here we pay attention to the small blob of galaxies that clusters apart from the main cloud in Fig. \ref{Fig:Pcurve4DturningPoints}. Another interesting task is to isolate a $pure$ population of red galaxies in the $L_{20}$ group whose luminosity function is Log-normal, as it appears in Fig. \ref{Fig:LumFun20sep}. 

\begin{figure}
\begin{center}
\epsscale{1.0}
\plottwo{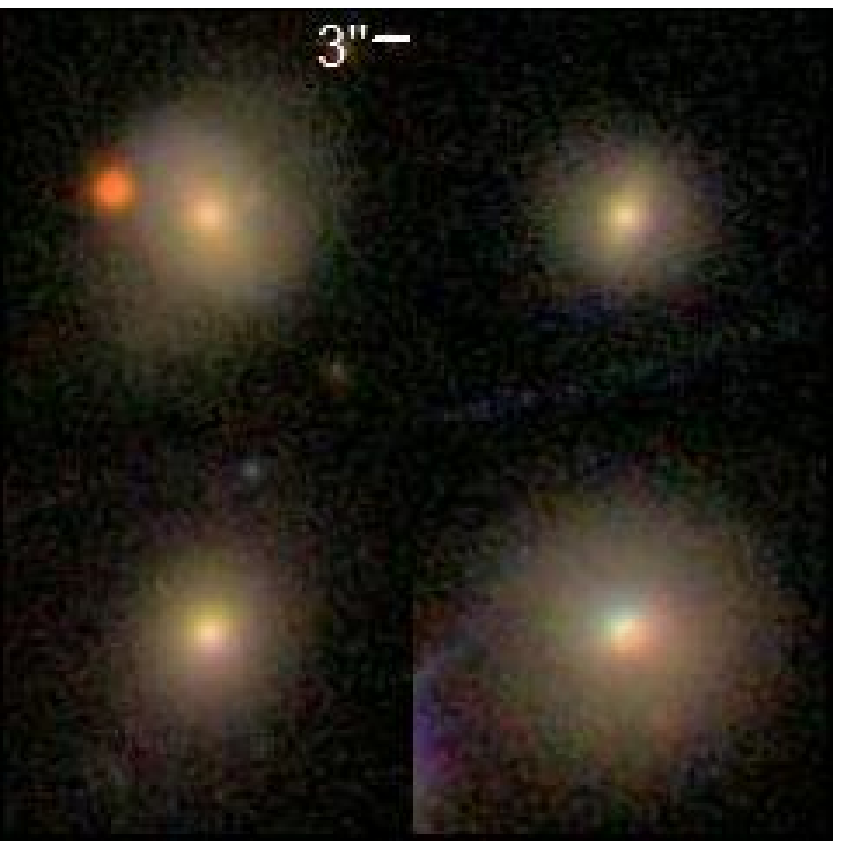}{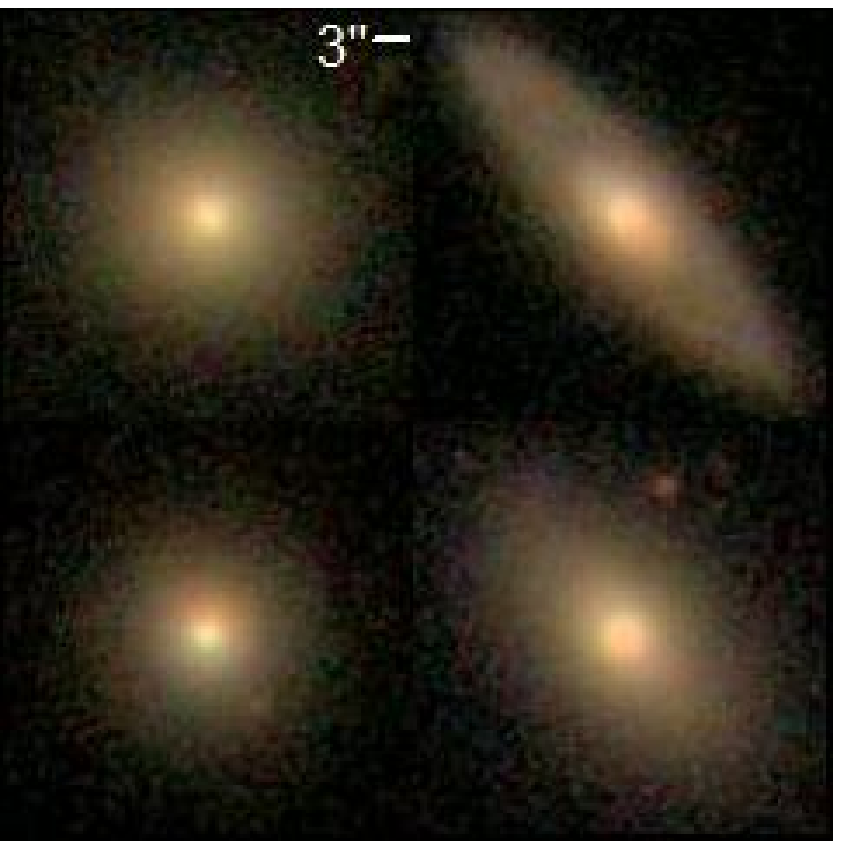}
\epsscale{2.2}
\plottwo{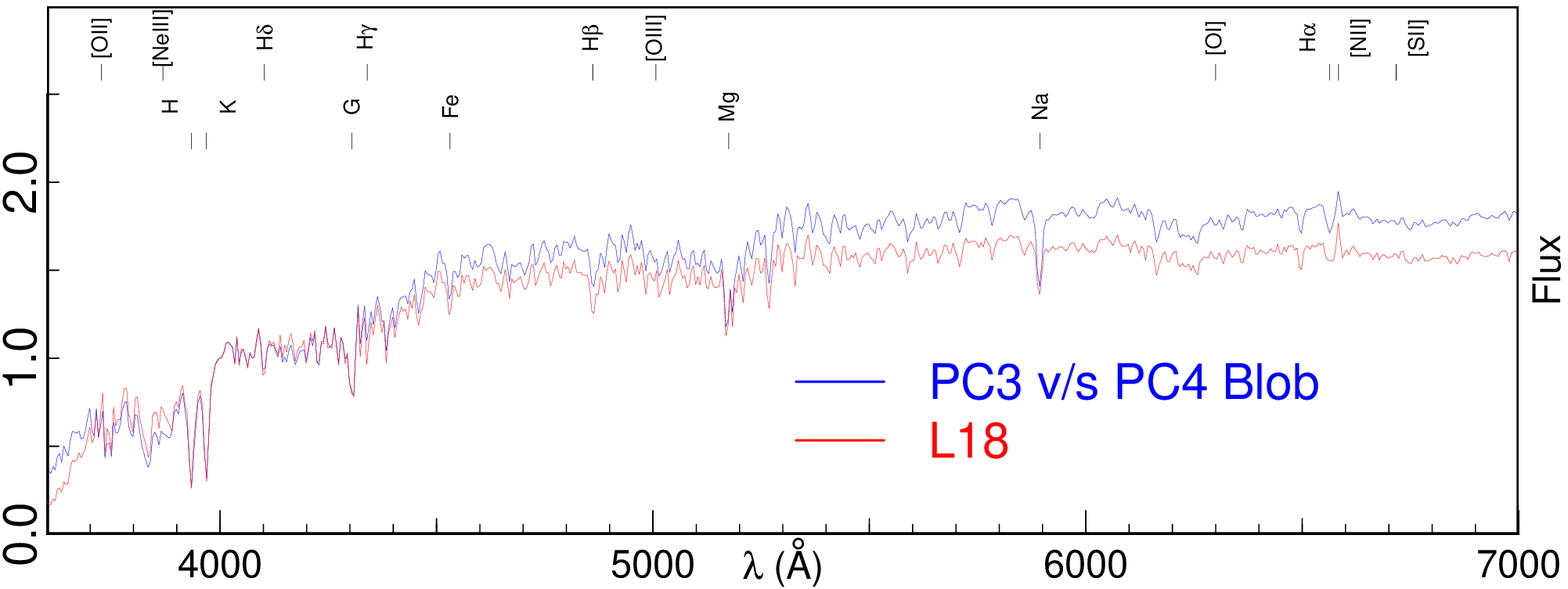}{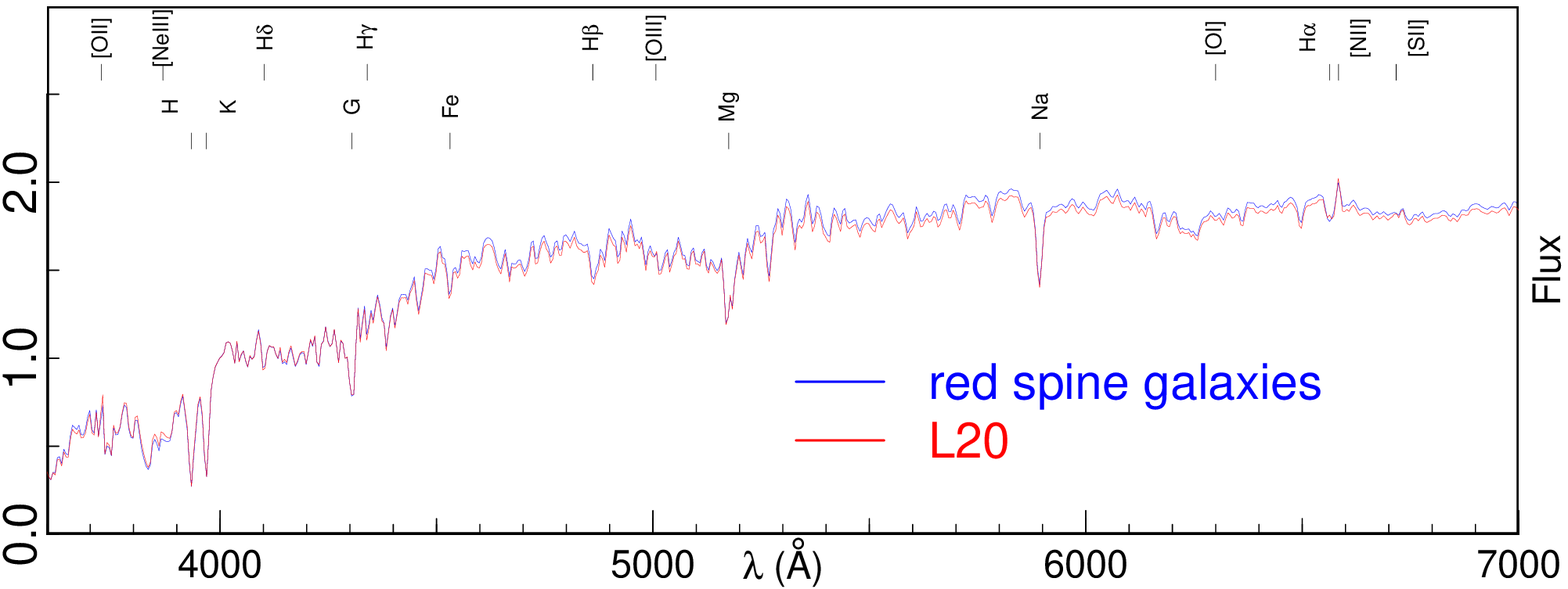}
\caption{Equivalent to Fig. \ref{Fig:MorphologyAndGalSpectra}, but showing interesting groups derived from WPCA and P-curve analysis. 
\\
\emph{Top}: Panels with the 4 most representative galaxy shapes for the small red disk-like galaxies blob (\emph{left}) and red spine galaxies (\emph{right}).
\\
\emph{Bottom}: Average spectrum for the 2 previous groups.}
\label{Fig:InterstingGroups} 
\end{center}
\end{figure}

\subsubsection{Blob of small red disk galaxies/lenticulars of high $M_{*}/L_{r}$ and $\mu_{*,50}$   }\label{Sec:RedSpiralsPC3PC4}

In Fig. \ref{Fig:Pcurve4DturningPoints} we found a small blob of galaxies clustered apart from the main cloud in the $\mathbf{PC}_{3}$ v/s $\mathbf{PC}_{4}$ panel. We separated them by using the following   separating line (built by eye): $\mathbf{PC}_{4} \leq -1.3 + 0.55(\mathbf{PC}_{3}-2.0)$. We further constrained these galaxies by choosing an appropriate interval in arc length $[l_{min},l_{max}]=[11.6,12.6]$, which includes most of the $L_{18}$ and part of $L_{19}$ groups. In fact, the blob appears also in Figure \ref{Fig:LambdaVsProp} bracketed within this arc length range in the $R90_{r}/R50_{r}$, $\mu_{*,r}$ and $\log R90_{r}$ panels. Precisely, $R90_{r}/R50_{r}$ is the property that dominates in $\mathbf{PC}_{3}$ and $\mathbf{PC}_{4}$ with opposite signs, as shown in Fig. \ref{Fig:PCAEigenvectors}. After the selection cuts, we are left with 136 blob galaxies, whose imaging and average spectrum are shown in Fig. \ref{Fig:InterstingGroups}.

According to Table \ref{Table:20GroupAveProp}, the blob is composed mostly by small red disk galaxies, with minimal star formation and void of gas. In fact, $u-r$ is at least 1$\sigma$ redder than the average in the $L_{18}$ group. Furthermore, they are modeled with an important component of an exponential disk ({\tt fracDeV}$_r =0.00^{+0.48}_{-0.00}$). The concentration index $R90_{r}/R50_{r} \simeq 1.95$ is low and far from the often used $R90_{r}/R50_{r}$ = 2.6 separator between ellipticals and spiral galaxies \citep{strateva2001}. The average size of $R90_{r}\simeq 1.6$kpc is small compared to the one of $L_{18}$ ($R90_{r}\simeq 4.0$kpc), lying beyond the 1$\sigma$ significance level. Interestingly, the small size makes $M_{*}/{\rm L}_{r} \simeq 4.07 M_{\odot}L^{-1}_{\odot,r}$, $\mu_{50,r} \simeq 18.83$ mag arcsec$^2$ and $\mu_{*,r}= 10^{10.06} M_{\odot}{\rm kpc}^{-2}$ to be also well above 1$\sigma$ of their average at $L_{18}$. With respect to the uncertainty of these values, errors in the estimate of $M_{*}$ can arise when fitting the spectra, specially when there is a strong component of dust, which these galaxies appear to have. The statistical error in $M_{*}$ is around 15\%. When comparing to a catalog of groups from \cite{tago2010}, we found that at least half of these galaxies are in groups of more than 10 members. We speculate that they were depleted of gas by ram pressure stripping or another mechanism that did not perturb the structure of the disk.

\subsubsection{Close-to-spine Red galaxies of Log-Normal luminosity function.}\label{Sec:RedSpineGaussianLumFun}

In Fig. \ref{Fig:LumFun20sep} we showed that the $L_{20}$ group of red galaxies has a luminosity function close to a log-normal distribution, clearly different to the Gamma and double power-law distributions of the other groups. For further study, we wanted to isolate the galaxies in this group whose luminosity function is exactly log-normal.

In order to extract these galaxies, we choose the ones falling in $\lbrace \lambda_{98}, \lambda_{99}, \lambda_{100} \rbrace$, which are the last 3 subgroups of $L_{20}$ as explained in \ref{Sec:PrincipalCurveAndPopSep}. We further chose the 1st partition closest to the P-curve, out of the 10 radial partitions in each $\lambda_{i}$ group, selecting therefore galaxies on or very close to the spine of the data point cloud.

The selected red spine galaxies are part of the very high density core of the red sequence of galaxies found in $L_{20}$. In fact, Table \ref{Table:20GroupAveProp} and Fig. \ref{Fig:LambdaVsProp} show that they have very similar properties values as their averages of the whole $L_{20}$. They are mostly red ellipticals ($u-r \simeq 2.62$, {\tt fracDeV}$_{r}\sim 1$ and $R90_{r}/R50_{r} \simeq$ 3.23), of mass $M_{*} \simeq 7.08 \times 10^{10}M_{\odot}$ and luminosity L$_{r} \simeq 2.86 \times 10^{10}$L$_{\odot,r}$. Figure \ref{Fig:LumFun20sep} shows that the luminosity function of is in fact very close to log-normal, with parameters shown in Table \ref{Table:FittingParametersLumFun}.


\section{Discussion}\label{Sec:Discussion}

The unsupervised non parametric methods of WPCA and P-curve should not be considered useful only for dimensionality reduction and easy data visualization. In this paper, these methods proved also being able to provide supporting evidence for some physical models and scenarios relevant in extragalactic astronomy, as discussed next.

\subsection{Information Content of the Principal Curve}\label{Sec:InformationContentPcurve}

The principal curve provides an objective way for ordering galaxies along its arc length. The success in dimensionality reduction and classification power of the P-curve is related to how much the projections of the galaxies are spread along arc length. In fact, a big variance along arc length gives more room for building separators and discerning different galaxy types. In our case, the arc length values along the P-curve have a variance of $\sigma^{2}_{l}=7.79$ (see Table \ref{Table:20GroupStatistics2}), bigger than cumulative variance $\Sigma_{i=1}^{i=4} \sigma^{2}_{\mathbf{PC_{i}}}=6.81$ of the principal components it was built from. This means that the curvature of the P-curve helps to discern information not included in the intrinsic linearity of WPCA. The length of the P-curve cannot be made arbitrarily long or short, due to an evident bias-variance trade-off. The shortest curves (with no curvature) are identical to $\mathbf{PC}_{1}$, with an average spread across it equal to $\Sigma_{i=2}^{i=p}\sigma^2_{\mathbf{PC}_{i}}$, and a high bias due to the straight P-curve missing the important bends in the structure of the cloud of points. On the other hand, the longest curves possible would be the ones connecting all the data points, which produce a null bias but high variance, as the curve will fit the noise in the structure of the cloud. In fact, as we experimented with values of $df \simeq 7$, the curve attempts to cover all the space spanned by the cloud, twisting and coiling itself in ways that describe additional detailed features of galaxies, while we are now interested in the global trends. Our election of $df=5.4$ is an intermediate case, where the root mean square of the projection distances on the curve is $d_{\perp}=1.31$, smaller than $(\Sigma_{i=2}^{i=4}\sigma^2_{\mathbf{PC}_{i}})^{1/2}=1.52$. The ratio 1-$\sigma^{2}_{l}/d^2_{\perp}=0.75$ gives us a notion of the amount of information that the P-curve is able to discern. The physical origin in the scatter of the remaining 25\% is still to be explored, and depends locally on the direction in the eigenspace along which $d_{\perp}$ is measured.

\subsection{Explaining the Zoo of Galaxies }\label{Sec:ExplainingZooOfGalaxies}

In our analysis, the P-curve has been able to recover the well known bimodality between the blue and red populations. Since galaxy properties are highly correlated, only a few properties should be enough for explaining the variations in the zoo of galaxies, namely $u-r$ (from $\mathbf{PC}_1$), $SFR$ (from $\mathbf{PC}_2$) and less importantly $R90_{r}/R50_{r}$ (from $\mathbf{PC}_3$). In fact, the variations recovered by the P-curve and its "W" shape depend strongly on $SFR$. The color $u-r$, almost linearly correlated with arc length, tracks specifically the 4000\AA break in the continuum, which gives a measure of stellar age and separates the early to late galaxy types. However, this is not enough, as $R90_{r}/R50_{r}$ tells about morphology, whereas the $SFR$ tracks the amount of material produced in recent star bursts, which shows as the strength of emission lines such as the Balmer series, and eventually correlates with the galaxy size, mass and luminosity. For example, within the blue population (bluer than the green maximum) we can find the low star-formation and surface brightness spiral galaxies separating the bluest star-forming spheroidals/irregulars and the redder star-forming spirals with a prominent bulge. The importance of the color and emission lines from star formation in explaining variations in galaxy populations has also appeared in previous PCA studies \citep[e.g.][]{yip2004b,coppa2011,gyory2011}.

\subsection{Additional Evidence supporting some Physical Models}\label{Sec:PhysicalResults}

\subsubsection{AGN Activity and Star Formation Quenching}\label{Sec:GreenMaxDiscussion}

The P-curve presents a green density maximum, between the blue and red ones. The green maximum shows an interesting feature in $\mathbf{PC}_{1}$ (Fig. \ref{Fig:PCAvsLambda}), colors and $M_{*}/L_{r}$ as a function of arc length (Fig \ref{Fig:LambdaVsProp}). The average of these properties keep on increasing as the arc length increases, except at the green maximum in $L_{16}$, where they stay constant or even decrease their values. This behavior, however, is not seen for example in $SFR/M_{*}$, whose average continues decreasing monotonically at $L_{16}$.
Note that $L_{16}$ is the last group which shows significant star formation and/or emission lines (see Fig. \ref{Fig:MorphologyAndGalSpectra}). Indeed, the equivalent width of $H_{\alpha}$ drops by 1 dex (see Table \ref{Table:20GroupStatistics2}) when moving to $L_{17}$, this last group is consistent with no $H_{\alpha}$ emission given the 1$\AA$ resolution of SDSS spectra. Furthermore, $L_{16}$ is the last group in the pure star forming region branch on the BPT diagram (Fig. \ref{Fig:BPTdiagram}), right before the bordering region separating the pure star forming and composite regions. Thus, higher arc length groups have basically small to null star formation activity and contain AGN. This is in agreement with the findings that AGN activity might be the cause for the shutdown of star formation in these galaxies \citep{martin2007}.

\subsubsection{Hierarchical Model of Galaxy Formation}\label{Sec:LogNormalLumFunDiscussion}

The luminosity functions shown in Fig. \ref{Fig:LumFun20sep} can be classified into roughly gamma and log-normal distributions. It has been shown, e.g. \cite{cooray2005} and \cite{yang2009}, that the LF of the Schechter fit for LFs can be divided into several components, coming from 2 different populations in dark matter halos: central or brightest cluster galaxies (BCGs) and satellite galaxies. Satellite galaxies are often given a power law LF, with a finite cut at the bright end given by the luminosities of the central galaxies. The centrals, on the other hand, are given bell-shaped luminosity functions. In particular, high mass halos ($M_{h}>10^{13}M_{\odot}$, according to \cite{cooray2005}), contain central galaxies whose luminosity functions can be modeled as a log-normal distribution. This is exactly the behavior observed for the $L_{19}$ and $L_{20}$ groups in Fig. \ref{Fig:LumFun20sep}, and better seen for the red spine galaxies shown in Sec. \ref{Sec:RedSpineGaussianLumFun}. Note that $L_{19}$ and $L_{20}$ corresponds to $L_{w}8$ in Fig. \ref{Fig:CorrFuncions}, which appears having an autocorrelation function with stronger power at $r_{p}=$10Mpc than any other group. On the other hand, at $r_{p} \lesssim 0.1$Mpc there is a clear loss of power. This is consistent with $L_{19}$ and $L_{20}$ being mostly composed by central galaxies. Note that there is still power in the autocorrelation function of $L_{w}8$ at $r_p<$ 0.1Mpc, which shows that there are some satellite galaxies (mostly red) in this sample, which is consistent with the faint end tail of the LFs in Fig. \ref{Fig:LumFun20sep}.

Log-normal distributions appear in nature as a consequence of multiplicative processes \citep[][and references therein]{lempert2001,mitzenmacher2003}, where the initial value $Y_{0}$ of a random variable is changed in successive steps in the form $Y_{j}=F_{j}Y_{j-1}$ by i.i.d multiplicative factors $F_{j}$ of distribution P(F). Using the central limit theorem for $j\rightarrow \infty$, it can be shown that $Y$ follows a log-normal distribution, independent of P(F). This argument can be extended to explain the log-normal luminosity functions of central galaxies and their stellar mass functions as well, since the r-band luminosity traces a population of old stars, which form the bulk of the mass of galaxies \citep{bell2003}. In fact, hierarchical galaxy formation models \citep[e.g][]{steinmetz2002,delucia2007} explain the creation of massive elliptical BCGs as a series of dry mergers of existing galaxies. Thus, a dense environment will allow several steps of mass adding or stripping that might lead to the formation of BCGs and cause the log-normal mass distributions for them.
\\


\acknowledgments

MTP acknowledges the use of the VO spectrum service for averaging galaxy spectra \citep{dobos2004}, and thanks Ching-Wa Yip, Mark Neyrinck, Timothy Heckman and Sean Moran for useful discussion.
The authors thank the anonymous referee for advise directed to enhance the impact of this paper.

Funding for SDSS-III has been provided by the Alfred P. Sloan Foundation, the Participating Institutions, the National Science Foundation, and the U.S. Department of Energy Office of Science. The SDSS-III web site is http://www.sdss3.org/.

SDSS-III is managed by the Astrophysical Research Consortium for the Participating Institutions of the SDSS-III Collaboration including the University of Arizona, the Brazilian Participation Group, Brookhaven National Laboratory, University of Cambridge, Carnegie Mellon University, University of Florida, the French Participation Group, the German Participation Group, Harvard University, the Instituto de Astrofisica de Canarias, the Michigan State/Notre Dame/JINA Participation Group, Johns Hopkins University, Lawrence Berkeley National Laboratory, Max Planck Institute for Astrophysics, New Mexico State University, New York University, Ohio State University, Pennsylvania State University, University of Portsmouth, Princeton University, the Spanish Participation Group, University of Tokyo, University of Utah, Vanderbilt University, University of Virginia, University of Washington, and Yale University.
 




\end{document}